\documentclass{aa}
\usepackage{graphicx}
\usepackage{txfonts}
\usepackage{natbib}
\bibpunct{(}{)}{;}{a}{}{,}
\usepackage{lscape}
\usepackage{rotating}
\usepackage[colorlinks,linkcolor=blue,anchorcolor=blue,citecolor=blue,]{hyperref}
\usepackage[switch]{lineno}

\newcommand{\hii}{\mbox{H\,\textsc{ii}}}
\newcommand{\hi}{\mbox{H\,\textsc{i}}}

\begin{document}

\title{Parallax-based Distances to Galactic $\hii$ Regions: Nearby Spiral Structure} \author{X. J. Shen\inst{1},  L. G. Hou\inst{2}, H. L. Liu\inst{3}, X. Y. Gao\inst{2}
}

\offprints{L.G. Hou, H. L. Liu\\ \email{lghou@nao.cas.cn,hongliliu2012@gmail.com}}

\institute{Department of Astronomy, School of Physics and Astronomy, Yunnan University, Kunming 650091, China \and National Astronomical Observatories, CAS, Jia-20 Datun Road, Chaoyang District, Beijing 100101, PR China \and School of Physics and Astronomy, Yunnan University, Kunming, 650091, P. R. China }

\date{Received; accepted}

\abstract 
{The spiral structure of the Milky Way is not conclusive, even for the disc regions in the solar neighbourhood. Especially, the arm-like structures uncovered from the over-density maps of evolved stars are inconsistent with the commonly adopted spiral arm models based on young objects.}
{We aim to re-examine the arm segments traced by young objects and better understand the nearby spiral structure.}
{We identify the exciting stars of 459 $\hii$ regions and calculate their parallax-based distances according to the {\it Gaia} DR3. Together with other $\hii$ regions with spectrophotometric or parallax-based distances in literature, the largest ever sample of 572 $\hii$ regions with accurate distances is used to reveal the features shown in their distributions projected onto the Galactic disc. The results are then compared to the features traced by other young objects (high-mass star-forming region masers, O-type stars, and young open clusters) and evolved stars. }
{The structures outlined by different kinds of young objects do not exhibit significant deviation from each other. The distributions of young objects are in agreement with three arm-like features emerging in the over-density map of evolved stars. Especially, the Local Arm outlined by young objects follows an arm-like feature delineated by evolved stars and probably spirals outwards towards the direction of $\ell \sim 240^\circ$ in the third Galactic quadrant. }
{We conclude that the arm segments traced by young objects and evolved stars are consistent with each other at least in the solar neighbourhood. In particular, 
the Local Arm delineated by young objects is reinterpreted as an arm segment with a large pitch angle of $25.2^\circ \pm 2.0^\circ$, whose inner edge is in good agreement with the recently discovered Radcliffe Wave.}


\titlerunning{Parallax-based distances to Galactic $\hii$ regions and nearby spiral structure}
\authorrunning{X.J. Shen et al.}

\maketitle
\section{Introduction}
\label{sec:intro}
The spiral structure of the Milky Way, its formation, and its evolution are longstanding problems. 
Although many efforts have been dedicated to uncovering the truth since the 1950s \citep[e.g.,][]{oort52,mwc53,vmo54,gg76,bg78,ch87,dcg+87,rus03,hhs09,reid14,reid19,xhb+21,xu+23}, the morphology of Galactic spiral arms, their accurate locations, and the kinematic properties remain elusive~\citep[see 
the reviews by e.g.,][]{fc10,xhw18,sz20,hou21}. 
The controversies are not confined to the distant Galactic disc regions but also exist in the solar neighbourhood for the spiral structure traced by more evolved stars and young objects.

In the last decade or so, the nearby spiral structure traced by young objects has been updated with the developments of the Very Long Baseline Interferometry (VLBI) technique~\citep[e.g.,][]{xrzm06} and the data releases of the {\it Gaia} mission~\citep{gaia18,dr323}.
Different kinds of young objects 
yield similar results.
The distribution of high-mass star-forming region (HMSFR) masers with VLBI parallax measurements has delineated three spiral arm segments (the Perseus Arm, the Local Arm, and the Sagittarius-Carina Arm) within about 5~kpc of the Sun, as well as a few spurs between them~\citep{xlr+13,xu+16,reid14,reid19}. 
The depicted arm segments were then confirmed by using a large number of massive O- and/or B-type stars with {\it Gaia} astrometric measurements~\citep[e.g.,][]{xu+2018,chh+19,pmb+21}, and extended to the southern hemisphere based on the all-sky data of the {\it Gaia} DR3~\citep[][]{xhb+21,xu+23}. 
The young open clusters \citep[YOCs, e.g., with ages $< 20$ Myr,][]{hxh21} are too young to migrate far from their birthplaces. Hence, they are excellent tracers of spiral structure. 
The distribution of {\it Gaia} YOCs~\citep[e.g.,][]{cjv18,scr18,lp19,hxw20,cac20,dmm21,hxw22} traces nearby spiral features consistent with that revealed by HMSFR masers and OB stars \citep[e.g.,][]{hxh21,jm23}. 
Similar analyses were also made by using young stellar objects by \citet{kbz+21}.

However, the situation gets complicated when evolved stars are used to trace the nearby spiral structure.
\citet{msk+19} analysed the surface over-density map of a sample of turn-off stars with typical ages of about 1~Gyr. They found an arm-like over density (see the definition in Sect.~\ref{stellararm}) possibly corresponding to the stellar Local Arm, which is close to the Local Arm traced by HMSFRs \citep{xu+2018}, but with a visible offset in position and a larger pitch angle.
Later, \citet{pdc+21} derived the over-density maps and displayed three nearby arm segments based on a sample of upper main sequence stars from {\it Gaia} DR3. The geometry of their suggested Perseus Arm and Local Arm differs significantly from that of many previous models based on young objects \citep[e.g.,][]{reid19}.
The Local Arm traced by upper main sequence stars extends into the third Galactic quadrant towards $l \sim 240^\circ$ as that of \citet{msk+19}.
A similar result about the stellar Local Arm was obtained by \citet{lxh22} by using a sample of Red Clump Stars from {\it Gaia} DR3, whose typical age is about 2~Gyr.
In addition, many efforts have also been made to depict the nearby spiral structure with stars and characterize the dynamic features~\citep[e.g.,][]{mgf15,kgd+20,mpp22,llk+22,jm23,amf+23,ugs23,gaiastruc23,dfa+24,wn24,glh+24}.
The different spiral patterns indicated by young objects and evolved stars pose an observational challenge to the scenarios of nearby spiral arms and the formation mechanism.
%

Therefore, many efforts in observations will be indispensable to enlarge the sample size of young objects and evolved stars with accurate measurements of basic parameters (distances, proper motions, velocities, etc.), which is the cornerstone of building an accurate map of the nearby spiral structure.
An accurate map of the spiral arms will be vital to test the possible formation mechanisms~\citep[][]{lin+64,tt72,sc84} of the Galactic spiral structure. Additionally, a few studies \citep[e.g.,][]{val14,hh15,vall22} suggest that the possible offsets are $\sim100-500$~pc near the arm tangent regions, however, such observational tests have not been well extended to other Galactic disc regions. In theory, the spatial offsets between the spiral arms indicated by different age populations arise only within the framework of the quasi-stationary density wave model \citep[e.g.,][]{db14}, which also requires further observational constraints.


$\hii$ regions, as a type of young objects are ionised gas clouds surrounding massive young stars or star clusters, with a typical electron temperature $T_e \sim$~8000~K and an electron density $n_e \sim$ $10^{2}-10^{4}$~cm$^{-3}$ \citep[][]{leq05}. 
Only the massive stars with spectral types of B2 or earlier can produce sufficient ultraviolet photons to ionise the surrounding interstellar medium and form $\hii$ regions \citep[][]{abb+14,aaw+21}.
The typical age of $\hii$ regions is only $\sim$10~Myr.
As the indicators of massive star formation at the present epoch, $\hii$ regions have long been used as excellent tracers of the Galactic spiral structure~\citep[e.g.,][]{gg76,ch87,rus03,hhs09}.
Although more than 7\,000 $\hii$ regions or candidates have been identified \citep[e.g.,][]{aaw+21}, the distances for most of them have not been determined with high accuracy~\citep[][]{abb+14,hh14}, which is the main obstacle to a better delineating the Galactic spiral structure with $\hii$ regions.

Three different ways have been commonly used to determine the distances of $\hii$ regions. They are the trigonometric parallax observations towards the associated
maser spots \citep[e.g.,][]{xrzm06,honm12,reid14,reid19}, the spectra-photometric measurements of the exciting stars \citep[][]{rus03,mdf+11}, and the kinematic method~\citep[e.g.,][]{hh14}.
The kinematic distances are model-dependent and sometimes present considerable uncertainties, relying on the choice of the Galactic rotation curve, the solution of the kinematic distance ambiguity problem, and the possible deviation from the assumed non-circular rotation~\citep[][]{benjamin08,xhw18}. 
In comparison, the trigonometric parallax is believed to be the best method, which has provided accurate distances for dozens of Galactic $\hii$ regions by identifying the associated HMSFR masers.  
The spectra-photometric method is not as accurate as trigonometric parallax and can generate relatively accurate distances with uncertainties of about 20\% for more than 200 $\hii$ regions \citep[e.g.,][]{rus03,mdf+11,fb15}.

In the past decade, the number of identified OB stars has increased considerably \citep[e.g.,][]{reed03,skiff14,mdb+15,msa+16,mdn+17,lcl+19}, most of which have accurate astrometric measurements by the {\it Gaia} mission.
It provides a good opportunity to identify the exciting stars and obtain accurate parallax distances for many Galactic $\hii$ regions.
This kind of work has only been done for 47 $\hii$ regions by \cite{maa+22} using {\it Gaia} DR3 to study the gradients of chemical abundances in the Milky Way. 
A systematic analysis for a large number of $\hii$ regions is absent and will be valuable for some studies, such as the abundance gradients in the Milky Way and the chemical evolution~\citep[e.g.,][]{wwb83,dpcc00,egp+05,rfb+06,qrbb+06,mc10,wba+19}, the luminosity function of $\hii$ regions~\citep[e.g.,][]{sk89,ftd90,mw97}, and triggered star formation by the expansion of $\hii$ regions~\citep[e.g.,][]{kg16,LiuHL15,LiuHL16}.

In particular, it motivates us to carry out the work that pays attention to the property of the Galactic spiral structure.
This work is organised as follows. In Sect.\,2, we describe the archived data for $\hii$ regions and massive OB stars. 
The analysis of parallax-based distances to a sample of Galactic $\hii$ regions is given in Sect.\,3. 
The results are presented in Sect.\,4, including the nearby spiral structure traced by $\hii$ regions complemented with other young and evolved objects for comparison. 
Discussions and conclusions are in Sect.\,5 and Sect.\,6, respectively.

\begin{figure*}
  \centering \includegraphics[width=0.99\textwidth]{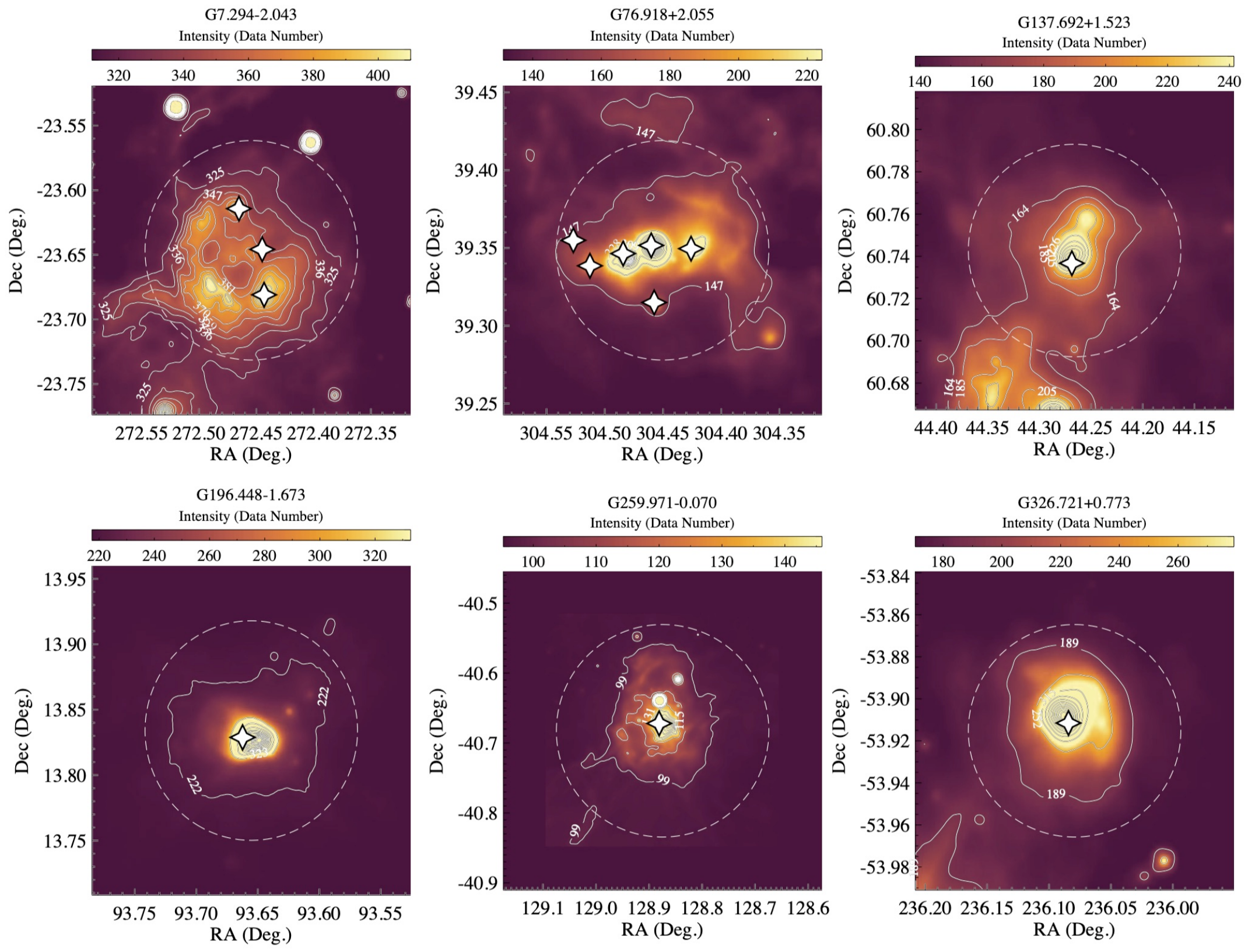} 
  \caption{Examples of $\hii$ regions with the candidate counterpart(s) of O-B2 stars (white star symbols). The {\it WISE} images of the W4 band (22\,$\mu$m, colour maps) are shown as backgrounds. The contour levels with the corresponding data numbers (white) are marked in each panel, and the first level is adopted to indicate the $\hii$ region boundary (see Sect.~\ref{sect:stars}). The dashed circle in each plot is the size of the $\hii$ region determined from the angular radius parameter given by the {\it WISE} $\hii$ region catalogue \citep[][]{abb+14}.}
\label{exampic}
\end{figure*}

\section{Data}

\subsection{Galactic $\hii$ regions}

The WISE~\citep[{\it Wide-field Infrared Survey Explorer},][]{wise} catalogue of $\hii$ regions \citep{abb+14} is the largest data set of Galactic $\hii$ regions and candidates. 
The catalogue lists about 8\,400 sources, including their coordinate, radius, recombination line velocity, and/or molecular line velocity if available from previous literature or latest observations \citep[e.g.,][]{abb+11,aaj+15,shrds21}. 
These sources were identified according to their mid-infrared
morphology. Essentially, all $\hii$ regions exhibit a good
mid-infrared morphology \citep{abb+14} in WISE observations:
the 12\,$\mu$m emission associated with an $\hii$ region primarily originates from polycyclic aromatic hydrocarbon molecules in the photodissociation region, 
surrounding the extended 22\,$\mu$m emission. 
The 22\,$\mu$m emission is mostly from hot dust, coincident with the
ionised gas traced by radio continuum emission \citep[e.g., see][]{dsa+10,abb+11} and/or recombination lines.

The massive young star(s) contributes to the ionising flux of an $\hii$ region and, hence, is expected to be close to the
mid-infrared 22\,$\mu$m, radio continuum, or recombination line emission peak(s).
For many Galactic $\hii$ regions, especially those with a radius smaller
than 1$^\prime$, there is a lack of high-quality radio continuum
images, partially because most of the existing radio continuum
surveys have a resolution larger than about 1$^\prime$
\citep[e.g.,][]{vgps,sgps,cgps}. An even worse situation is found for the available survey dataset of radio recombination lines \citep[e.g.,][]{hipass,siggma,gdigs,hhh22}.
In comparison, the infrared image data for Galactic $\hii$ regions always have better angular resolution. 
For the {\it WISE} 22\,$\mu$m dataset, its resolution
(12$^{\prime\prime}$) and sensitivity (6\,mJy) make it possible to
identify the detailed mid-infrared morphology of many Galactic $\hii$ regions.

In this work, we will identify the central exciting star(s) of
WISE $\hii$ regions by matching their 22\,$\mu$m morphology with the
known O- and early B-type stars. 

\subsection{Massive young stars}

In the past few decades, there have been considerable efforts to identify Galactic OB stars~\citep[e.g.,][]{reed03,mdb+15,msa+16,mdn+17,lcl+19}.
\citet{skiff14} has compiled a catalogue of OB stars from literature, which has been updated until 2022 May 13, and contains 138\,604 spectroscopically confirmed OB stars.
To our knowledge, it is the largest sample of spectroscopically confirmed OB stars available to date.
In addition, \citet{chh+19} identified 6\,858 highly confident
candidates of O- and early B-type stars by using the
photometric data of the VST Photometric H$\alpha$ Survey
\citep[][]{dgg+14} and the second {\it Gaia} data release \citep[][]{gaia18}. Together with 8\,022 known O-B2 stars in literature, they built a sample of O- and early B-type stars with robust parallax distances (parallax uncertainties $<$ 20\%) and also proper motions.

We adopt the catalogues of \citet{skiff14} and \citet{chh+19} as the sample of O-B2 stars to initiate this work. 
First, we check and solve the redundancy problem for these two catalogues. 755 duplicates are identified based on the specific spectral types as well as the uniqueness of the {\it Gaia} DR3 ID, and removed accordingly.
Then, we update the parallaxes and proper motions for the O-B2 stars with the {\it Gaia} DR3~\citep[][]{dr323}, which has significantly improved the accuracy of parallax measurements, with median parallax uncertainties of 0.02$-$0.03~mas for the {\it G} band magnitude $<$15, 0.07~mas at $G=$~17, 0.5 mas at $G=$~20, and 1.3 mas at $G=$~21 mag.

\begin{figure*}
  \centering
  \includegraphics[width=0.98\textwidth]{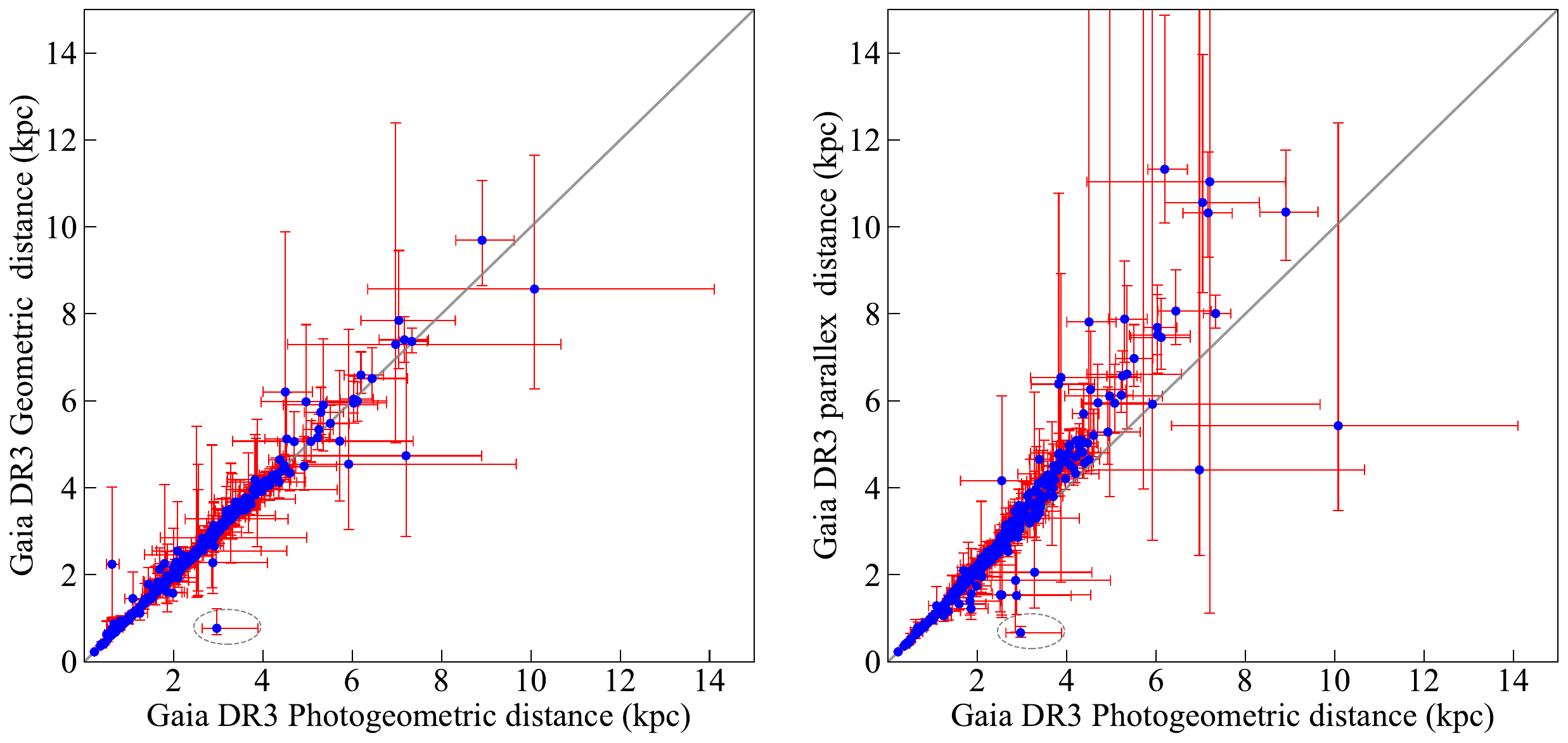}  
  \caption{Comparison of the photogeometric distances of the 459 $\hii$ regions with the other two types of distance determined from the {\it Gaia} DR3. {\it Left}: photogeometric distance vs. geometric distance. {\it Right}: photogeometric distance vs. inverse parallax. The grey line in each panel represents the equal line. The grey dashed ellipse denotes the $\hii$ region G331.172$-$00.460 (see Sect.~\ref{sect:dis}).}
\label{comparison}
\end{figure*}

\section{Distances of Galactic $\hii$ regions based on {\it Gaia}}
\subsection{Exciting stars}
\label{sect:stars}

To identify the exciting stars of $\hii$ regions, first of all, we single out the candidates by searching the O-B2 star(s) situated within the angular radius of an $\hii$ region.
781 $\hii$ regions are selected with 3\,211 O-B2 stars potentially associated.
Then, we inspect each of them by using the
SIMBAD database \citep[][]{simbad}, in case some O-B2 stars are not listed in \citet{skiff14} and \citet[][]{chh+19}, but given by the SIMBAD. 
283 additional O-B2 stars are obtained, which are confirmed spectrally with parallax measurements.
%
%
We hereto obtain 781 $\hii$ regions with 3\,494 candidate O-B2 stars (also see Table~\ref{summary}).

As mentioned in \citet{abb+14}, the angular radius of an $\hii$ region was determined based on the extent of the 12\,$\mu$m emission tracing the photodissociation region, which encloses the associated 22\,$\mu$m mid-infrared emission originating from warm dust within the ionised volume of the $\hii$ region. Considering that the 22\,$\mu$m morphology is generally complex (e.g., see Fig.~\ref{exampic}), the boundary of an $\hii$ region needs to be described as accurate as possible to count the exciting star(s).
First, we extract regions outside the $\hii$ region radius and free from bright 22\,$\mu$m emission.
The mean and standard deviation are calculated from these extracted pixels, respectively, as the background value and $\sigma$.
The $\hii$ region boundary is then defined as the 3$\sigma$ level above the background. Overall, the newly defined boundaries of the $\hii$ regions are,  as expected, smaller than that determined by \citet{abb+14}.
Accordingly, we exclude O-B2 stars for the related $\hii$ regions falling within their radius but outside their boundary.
We notice that tens of $\hii$ regions show large angular radii (e.g., more than hundreds of arcseconds) with very weak 22\,$\mu$m emission, they are discarded in this work since their boundaries are difficult to define with high confidence.  We have 473 $\hii$ regions with 1\,826 candidate O-B2 stars. Additional 14 $\hii$ regions are further
discarded as the parallax values of their candidate O-B2 stars are not consistent with each other in the 3$\sigma$ level. 
%
After these steps, we have 459 $\hii$ regions with 1\,592 candidate O-B2 stars (see Step\,3 in Table\,\ref{summary}). These O-B2 stars have parallax uncertainties at a median value of 4.6\%.

It is worth noting that despite our rather strict selection procedures, some of the candidate stars may be foreground/background sources. To alleviate this issue, we require that the parallax values of most stars falling within the boundary of an $\hii$ region should be consistent  within the 3$\sigma$ level. 
For 75 of the 459 $\hii$ regions, some candidate stars show a large difference beyond the 3$\sigma$ level from the major population, and are thus treated as foreground/background stars. 
Following this standard, about 19.5\% of the 1\,075 candidate stars within these 75 $\hii$ regions are eliminated. 
The mean radius of these 75 $\hii$ regions is $\sim$829$^{\prime\prime}$, larger than that of the 459 $\hii$ regions of $\sim$430$^{\prime\prime}$, favouring their possibility of being foreground/background stars. 
%

%
%

%

%

To sum up briefly, we pick out 459 $\hii$ regions with 1\,382 candidate exciting stars associated  (see Step\,4 in Table\,\ref{summary}).
Six exemplar $\hii$ regions are shown  
in Fig.~\ref{exampic}.
The parallaxes, proper motions, and their uncertainties from {\it Gaia} DR3 are given in Table\,\ref{exciting stars} for the stars, which will be used in the following to calculate the $\hii$ region distances.

\begin{table}[htbp]
\centering
\renewcommand\arraystretch{1.3}
\footnotesize
\caption{Summary of the number of sources retained at each step.}
\label{summary}
\vspace{1mm}
\tabcolsep10pt 
\begin{tabular}{lc}
\hline
\hline
Step ID & Number of Stars /{\bf $\hii$ regions} \\
\hline
Step-1   & 3494 / {\bf781} \\
Step-2   & 1826 / {\bf473}\\
Step-3   & 1592 / {\bf459}\\
Step-4   & 1382 / {\bf459}\\
\hline
\end{tabular}
\tablefoot{ The number of retained O-B2 stars and $\hii$ regions after each analysing step mentioned in Sect.~\ref{sect:stars}. "Step-1" identified O-B2 stars that fall within the angular radii of $\hii$ regions. "Step-2" was to select O-B2 stars located within the newly defined boundaries of $\hii$ regions. "Step-3" checked the parallax consistency of the stars within the boundaries of $\hii$ regions and the parallax uncertainties of stars. "Step-4" excluded the potential foreground/background stars whose parallaxes do not fall within 3$\sigma$ of that of the majority member stars.}
\end{table}

\begin{table*}[!t]
\centering
\renewcommand\arraystretch{1.3}
\footnotesize
\caption{$\hii$ regions with parallax distances based on {\it Gaia} DR3.}
\label{knownhii}
\vspace{1mm}
\tabcolsep10pt 
\begin{tabular*}{\textwidth}{lcllllclcl}
\hline
\hline
$\hii$ region Name & Cl & R &  $d_{\sun}$ & V$_{LSR}$ & Line & Ref. & d$_{kin}$ & Group & Note \\
& & ($^{\prime\prime}$)& (kpc) & (km~s$^{-1}$) & & &(kpc) & &\\
(1)&(2) & (3)& (4) & (5) &(6) & (7) &(8) &(9) &(10)\\
\hline
G000.003+00.127   & K & 206 & $2.28_{-0.12}^{+0.18}$& $-$5.5 & H$\alpha$ & 1 & -& I & -\\
G000.121$-$00.304 & Q & 37  & $2.51_{-0.11}^{+0.15}$& - & - & - &-& III & -\\
G011.662$-$01.692 & K & 907  & $1.25_{-0.02}^{+0.02}$ & 18.3 &
 RRL & 2 &-& I & maser\\
G017.315+00.389   & C & 78  & $1.85_{-0.05}^{+0.05}$ & - & - & - & - & III & -\\
G018.253$-$00.298 & K & 205 & $3.36_{-0.38}^{+0.33}$ & 47.5 & H$110\alpha$ & 3 & $3.60_{-0.36}^{+0.32}$ & I & M22\\
G018.669$-$00.298 & K & 272 & $1.95_{-0.06}^{+0.08}$ & 27.6 & CO(1$-$0) & 4 & $2.33_{-0.24}^{+0.17}$ & I & M22\\
G024.138+00.123   & K & 88  & $2.18_{-0.06}^{+0.08}$ & 114.5 & H$87+88/85\alpha$ & 2 & $6.61_{-0.32}^{+0.41}$ & III & -\\
G026.443+01.741   & K & 201 & $2.40_{-0.05}^{+0.07}$ & 42.3 & CS(2$-$1) & 5 & $2.70_{-0.22}^{+0.22}$ & I & M22\\
G026.802+03.526   & K & 69  & $0.41_{-0.01}^{+0.01}$ & 45.3 & H$\alpha$ & 6 & $2.88_{-0.27}^{+0.22}$ & III & -\\
G028.200$-$00.050 & K & 60  & $2.49_{-0.11}^{+0.14}$ & 96.8 & HCO$^+$ & 7 & $5.50_{-0.32}^{+0.40}$ & I & -\\
... & ... & ...  & ... & ... & ... & ... & ... & ... & ...\\
\hline
\end{tabular*}
\tablefoot{The entire version in a machine-readable format is available at the CDS. Column (1) is the source name. Cols.\,(2) and (3) are the classification and radius given by the WISE catalogue of $\hii$ regions \citep{abb+14}. Col.\,(4) is the parallax-based distance given by this work according to {\it Gaia} DR3 (see Sect.~\ref{sect:dis} for detail).  Cols.\,(5), (6), and (7) list the LSR velocity, the observed line(s), and the reference if available in literature: 1: \,\cite{ftd90}; 2:\,\cite{lockman89}; 3:\,\cite{dwbw80}; 4:\,\cite{bfs82}; 5:\,\cite{bnm96}; 6:\,\cite{ftd90}; 7:\,\cite{sck+04}. Col.\,(8) is the kinematic distance calculated by using the tool of \cite{wba+18} and the Galactic rotation model of \cite{reid19}. Col.\,(9) is a classification based on the robustness of the estimated parallax distances as discussed in Sect.~\ref{sect:dis}. Col.\,(10) gives a note for those $\hii$ regions with determined parallax distances in literature: maser: the distance is determined from the associated HMSFR maser; M22: the distance was given by \citet{maa+22} based on {\it Gaia} DR3.\\
}
\end{table*}

\subsection{Distances to $\hii$ regions}
\label{sect:dis}

Once the exciting star(s) have been identified, the distance of an $\hii$ region can be derived from the stellar distance(s).
To infer reliable distances for the vast majority of {\it Gaia} stars from the parallax measurements, a series of probabilistic methods have been developed \citep[e.g.,][]{bai15,ab16,ab16b,brf+18,chh+19}.
These methods generally involved a likelihood based on the measurements or the data processing, as well as a prior distribution that describes the space density distribution of stars. 
Based on {\it Gaia} DR3, \cite{brf+21} has estimated two types of distance for 1.47 billion stars, i.e., the geometric distance and the photogeometric distance. 
Both methods adopted a uniform direction-dependent prior which varies smoothly as a function of the Galactic longitude and latitude, considering the colours and magnitudes of the {\it Gaia} stars, as well as a zero-point offset for the parallaxes. 
Especially, the photogeometric distances utilized additional information of the {\it G}-band magnitudes and the BP-RP colours of the {\it Gaia} stars, which helps to improve the precision of distance estimates, especially for the stars in distant Galactic regions and the stars with poor parallax measurements. 
We cross-match our sample of exciting stars with the catalogue of \cite{brf+21} to obtain both types of distance. In addition, a distance directly estimated from the inverse parallax is considered for comparison.

Before choosing the most suitable distance estimator, we derive the distances of these $\hii$ regions based on the stellar data. 
For an $\hii$ region with only one associated star, the distance and its uncertainty are directly taken from the stellar data.
In cases where an $\hii$ region has at least two associated stars, 
we 
calculated the mean value of stellar distances, which is taken as the $\hii$ region distance. 
Based on the above criteria, the 459 $\hii$ regions are divided into three different groups as annotated in Table\,\ref{knownhii}.
(1) Group I: the $\hii$ region has more than one associated star and all the stars present consistent distances. Therefore, the estimated distances of these 160 $\hii$ regions in Group\,I ought to be reliable.
(2) Group II: the $\hii$ region originally has two or more associated stars, but only one is left after the above criteria.
(3) Group III: the $\hii$ region has only one associated star.
In total, Groups\,II and III contain 299 $\hii$ regions. We find their mean radius  to be $\sim$290$^{\prime\prime}$, 1.5 times less than that of the entire selected $\hii$ region sample of $\sim$430$^{\prime\prime}$. We therefore assume that most (if not all) of such $\hii$ regions have a good association with the corresponding single candidate star, and thus have a reliable distance determination.


The comparisons of different types of the $\hii$ region distance are shown in Fig.\,\ref{comparison}.
The photogeometric distances are consistent with the geometric ones considering the uncertainties for most $\hii$ regions.
An outlier G331.172$-$00.460 is near the Galactic plane, hence the correction for extinction may be insufficient \citep{brf+21,maa+22}, resulting in poor reliability of the photogeometric distance. 
As shown in the right panel of Fig.\,\ref{comparison}, the distances directly estimated from the inverse parallaxes tend to give larger values than the photogeometric or geometric method, especially for the sources with distances $\gtrsim$ 3~kpc.
In distant Galactic regions, the photogeometric distances typically have smaller uncertainties than the other two methods for most $\hii$ regions. 
Considering the suggestions given by \cite{brf+21} and \cite{gaiastruc23}, the photogeometric distance is preferentially adopted in this work and listed in Table\,\ref{knownhii} and Table\,\ref{exciting stars}.  
In addition, if the photogeometric distance is unavailable or in the case of G331.172$-$00.460, the distances are replaced with the geometric values and annotated in the Tables.

\subsection{Comparison with previous results}
\label{sect:com}

\begin{figure}
  \centering
  \includegraphics[width=0.47\textwidth]{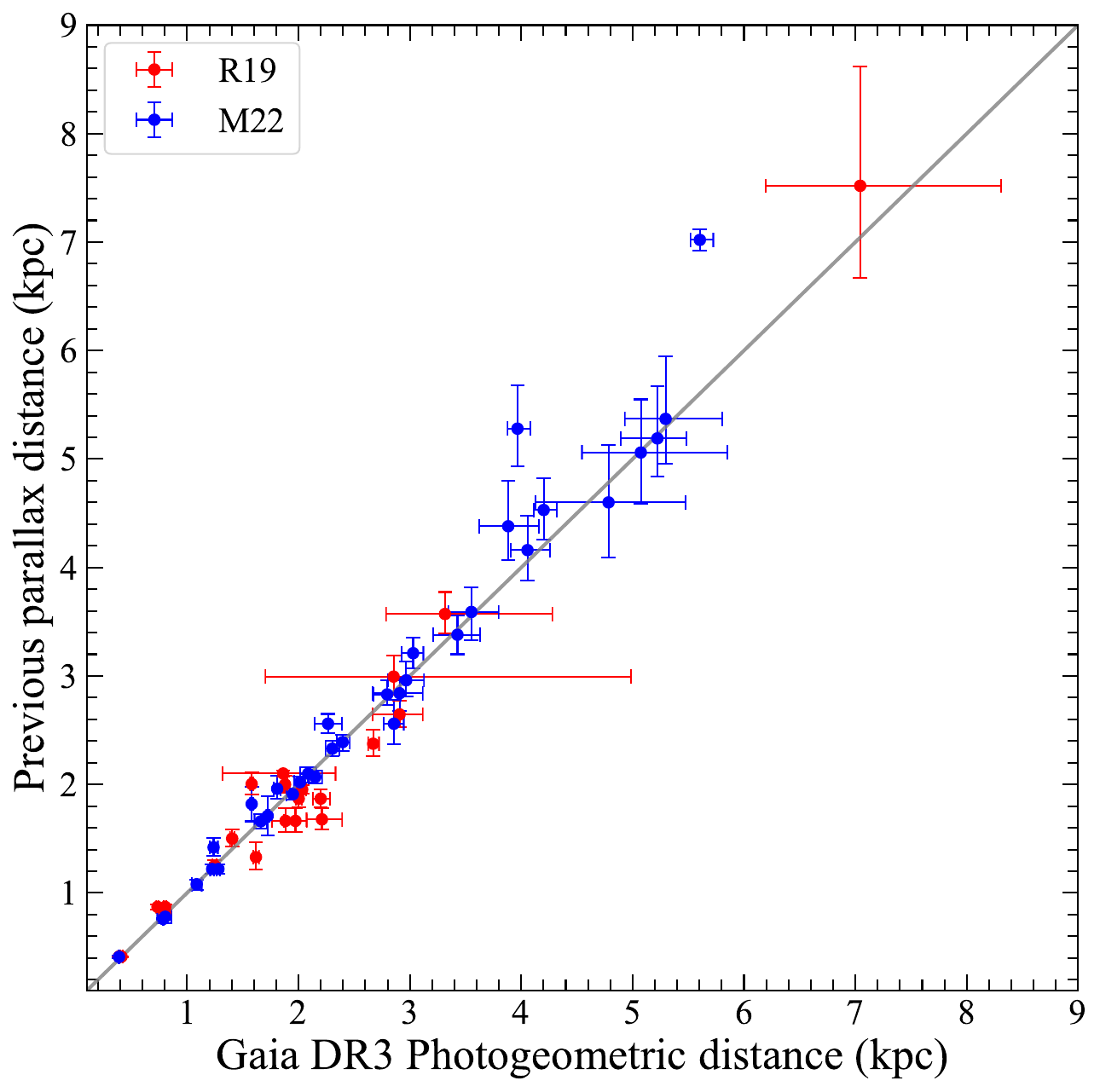}  
  \caption{Comparison between the parallax-based distances of $\hii$ regions given by this work and the reference values. The red dots indicate the sources with estimated distances from the associated HMSFR masers primarily from \cite{reid19}. The blue dots are the $\hii$ regions from \cite{maa+22}, whose distances are based on {\it Gaia} DR3. The grey line represents the equal line. Abbreviations for the references are: R19, \citet{reid19}; M22, \citet{maa+22}.} 
\label{distance}
\end{figure}

\begin{figure}
  \centering
  \includegraphics[width=0.47\textwidth]{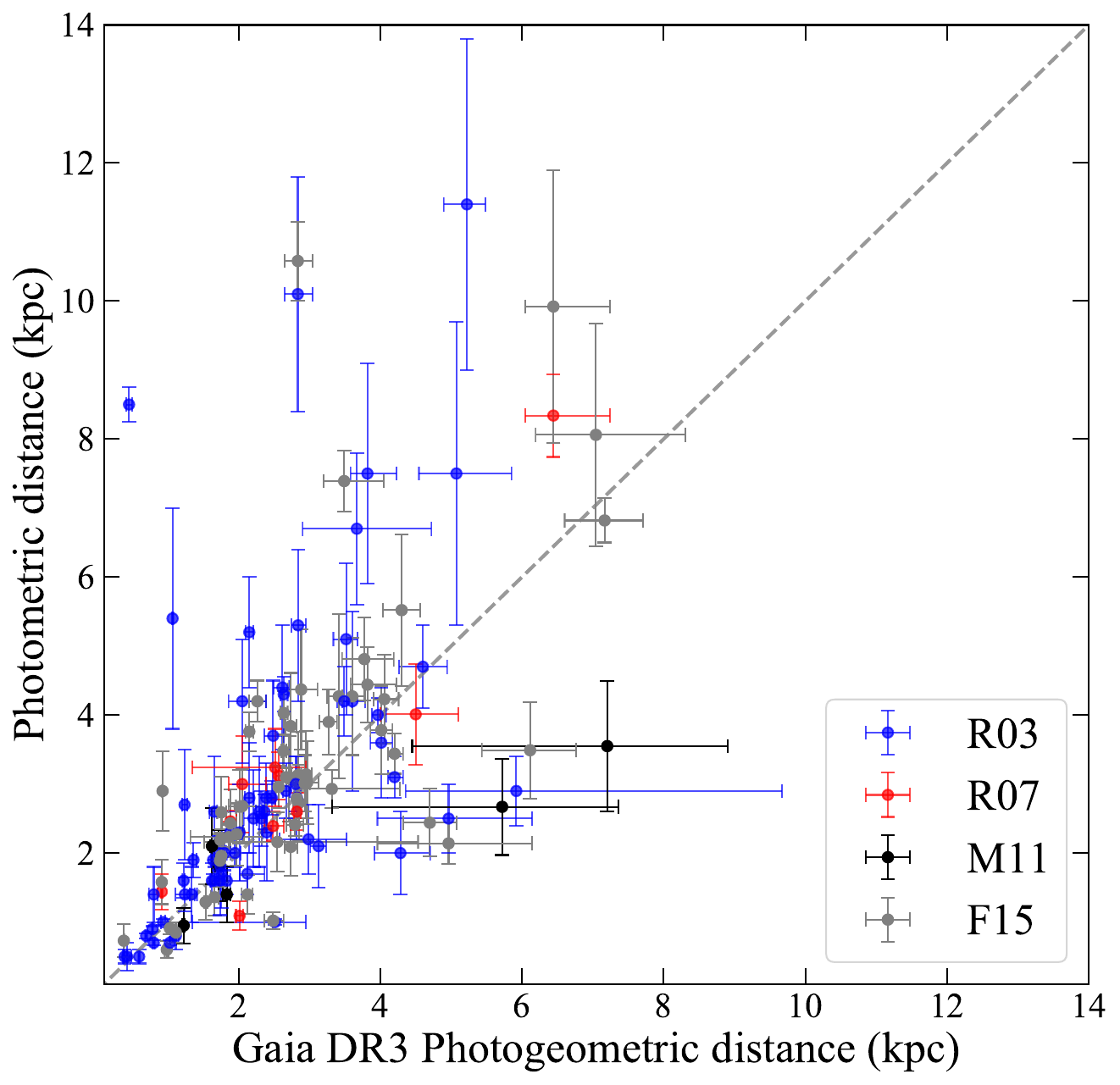}  
  \caption{Comparison between the parallax-based distances of $\hii$ regions given by this work and the spectrophotometric distances obtained from previous works. Abbreviations for the references are: R03, \citet{rus03}; R07, \citet{rag07}; M11, \citet{mdf+11}; F15, \citet{fb15}. The dashed grey line represents the equal line.}
\label{php}
\end{figure}

\begin{figure}
  \centering
  \includegraphics[width=0.48\textwidth]{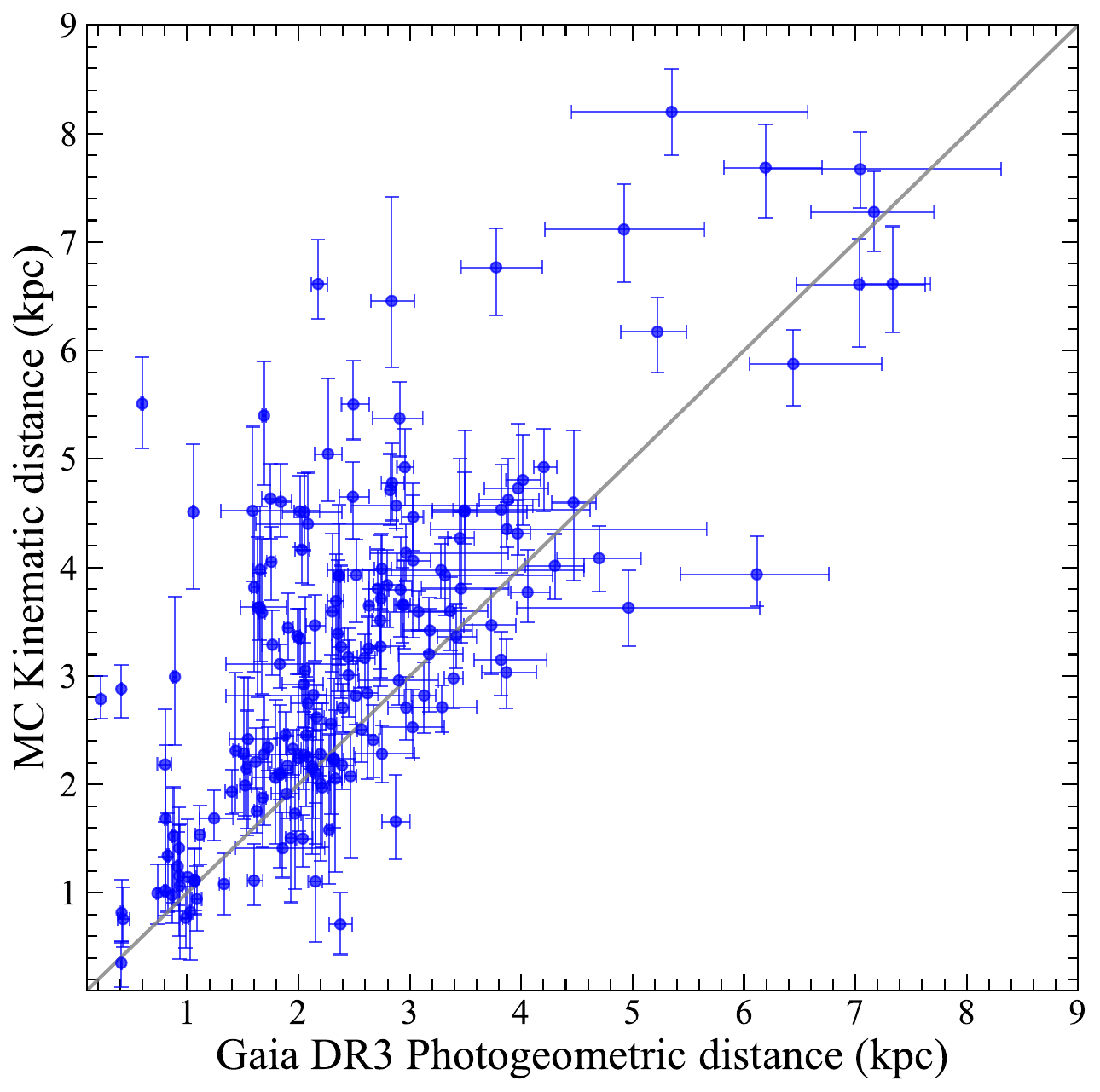} \\ \includegraphics[width=0.48\textwidth]{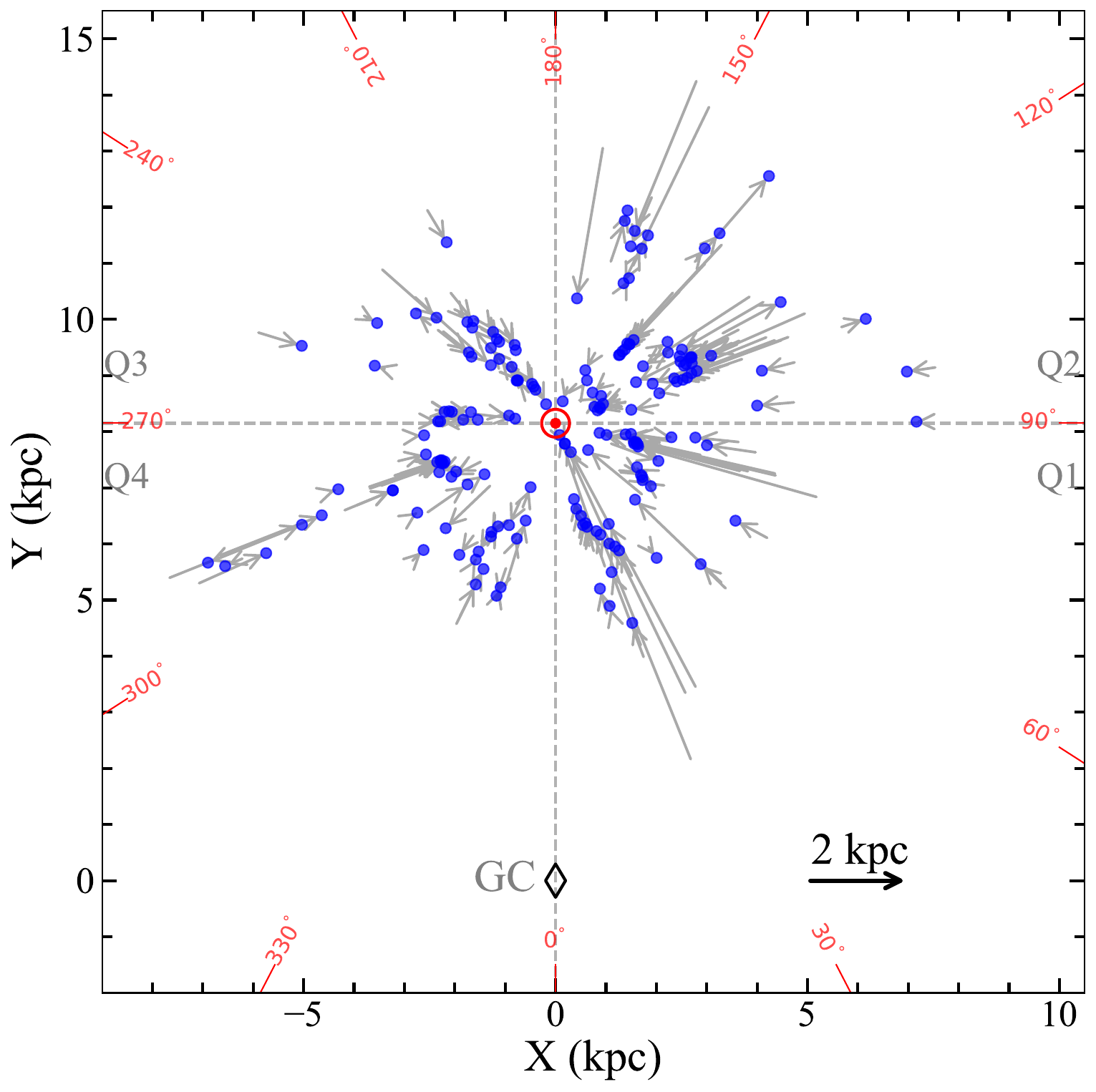}   
  \caption{{\it Upper}: comparison between the parallax-based distances of $\hii$ regions given by this work and the kinematic distances. The grey line is the equal line. 
  {\it Lower}: distributions of the $\hii$ regions projected onto the Galactic disc, to better show the influence of updated distance parameters on the delineated spiral arm segments.
  The starting point and ending point of each arrow represent the positions of an $\hii$ region according to the kinematic distance and the parallax-based distance, respectively. 
  The locations of $\hii$ regions based on the parallax distances are marked with blue dots. 
  In the plot, the Sun is at (0, 8.15) kpc \citep[][]{reid19}, while the Galactic Centre is at (0.0, 0.0) kpc. 
%
  }
\label{discom}
\end{figure}

Part of the 459 $\hii$ regions also have determined distances in literature, either by the trigonometric parallax method, the spectra-photometric measurements, or by the kinematic method. 
We compare the derived parallax-based distances in this work with the reference results.

The HMSFR masers with trigonometric parallax measurements are collected from the literature \citep[e.g.,][and references therein]{reid19,vera20,xu+23,bian+24} and cross-matched with the 459 $\hii$ regions.
29 HMSFR masers are falling within the $\hii$ regions defined by the angular radii~\citep[][]{abb+14}.
After an inspection of the mid-infrared images of the $\hii$ regions, five HMSFR masers were excluded because they are located outside the $\hii$ region boundaries (see Sect.~\ref{sect:stars}).
Then, the distances of the remaining 24 $\hii$ regions (see red dots in Fig.\,\ref{distance}) are calculated according to the parallax data of the associated HMSFR masers. 
As shown in Fig.~\ref{distance}, the results are in agreement with the parallax-based distances adopted in our work.
Meanwhile, \cite{maa+22} has estimated the distances for 47 $\hii$ regions by employing a method of identifying the ionising/associated stars located within the boundaries of $\hii$ regions based on {\it Gaia} DR3. 
The geometric distance of {\it Gaia} Early Data Release 3 or the kinematic distances of \citet{wba+19} are adopted to represent the distances for their sample.
Additionally, their procedures and criteria for identifying the associations are not the same as in this work.
We cross-matched our sample with their results and found that there are 34 $\hii$ regions overlapped. 
A comparison of their distance values (see blue dots in Fig.\,\ref{distance}) with our results is presented in Fig.\,\ref{distance}, which also demonstrates consistency for most of the targets.

\begin{figure*}
  \centering
  \includegraphics[width=0.9\textwidth]{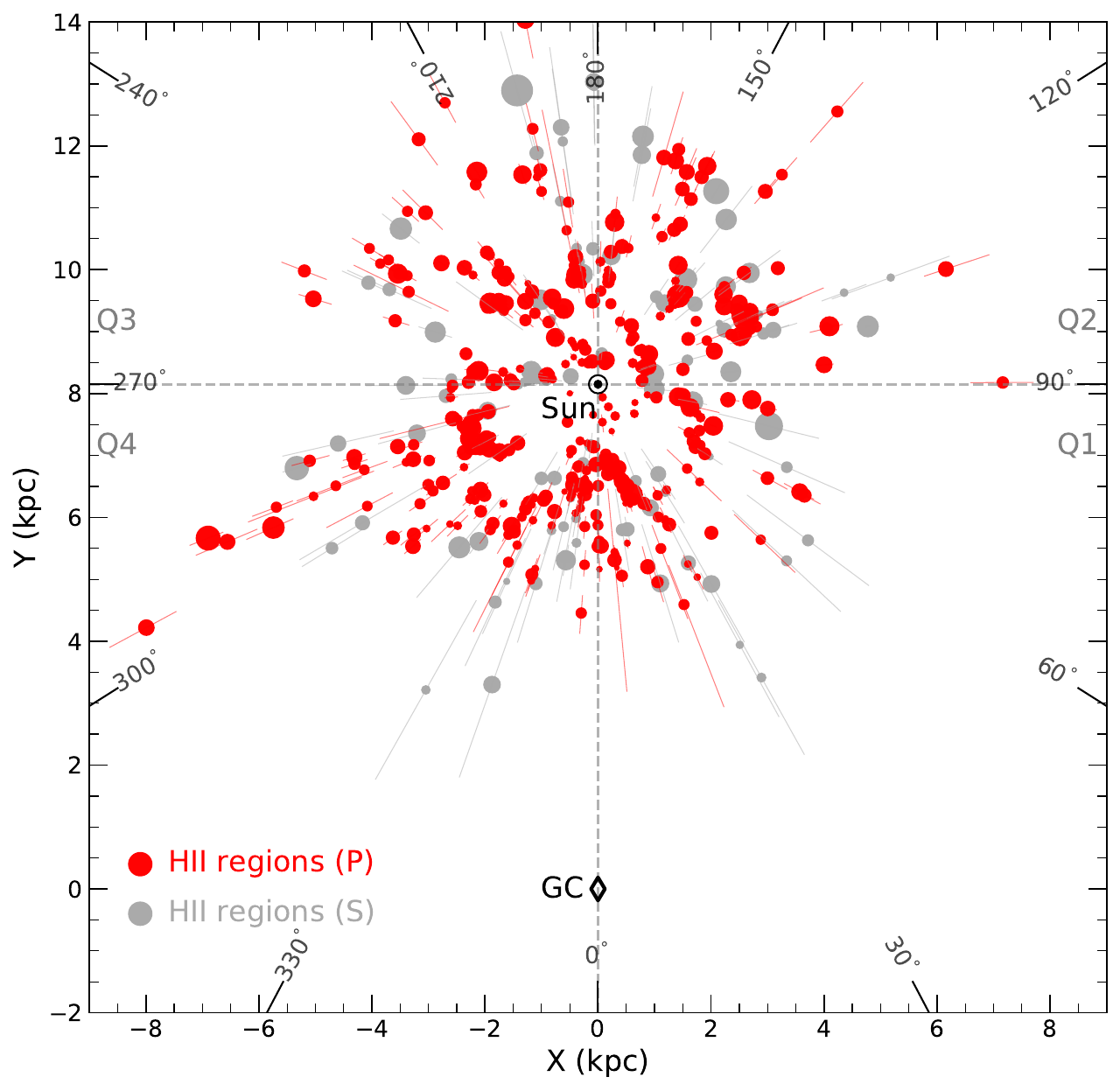}  
  \caption{Distributions of 572 $\hii$ regions (Sect.~\ref{sect:sample}) with accurate distances projected onto the Galactic disc. The $\hii$ regions with parallax-based (P) distances are shown by red dots, while those with spectrophotometric (S) distances are indicated by grey dots.  Also shown in the plot are the distance uncertainties (underlying line segments). The dot size is proportional to the estimated radio luminosity of the $\hii$ region at 5~GHz. The positions of the Sun and GC are the same as the lower panel of Fig.\,\ref{discom}. 
  }
\label{proj_hii}
\end{figure*}
The spectra-photometric method is not as accurate as the trigonometric parallax method but is still capable of providing relatively accurate distances.
We collected the $\hii$ regions with spectrophotometric distances determined in previous works \citep[][]{rus03,rag07,mdf+11,fb15}, and cross-matched the sample with our results.
Fig.\,\ref{php} presents a comparison between the spectrophotometric distances and our results for the 104 matched $\hii$ regions.
For most of the $\hii$ regions with distances $\lesssim$~4~kpc, these two methods give consistent results, although the spectrophotometric distances typically have larger uncertainties.
But in distant regions (e.g., $\gtrsim$~4\, kpc), these two distances tend to have larger deviations for most sources, which is probably related to the uncertainties on the Galactic extinction models.

At present, most of the known $\hii$ regions only have kinematic distances, which sometimes present large uncertainties.
With the information from the aforementioned literature, as well as the catalogues of \citet{shrds21} and \citet{siggma}, we collect the radial velocities $V_{\rm LSR}$ in the frame of the Local Standard of Rest (LSR) for 267 of 459 $\hii$ regions. 
These velocities were obtained by observing the ionised gas using radio recombination lines, optical H$\alpha$ line, and transitions from different molecular species (e.g.,$^{12}$CO, $^{13}$CO, CS, and NH$_3$). 
Then, the kinematic distances of these $\hii$ regions are re-calculated according to the Galactic rotation model of \cite{reid19} with the tool developed by \cite{wba+18}, which involved a Monte Carlo technique to account for the distance uncertainties. 
If the kinematic distance ambiguity problem has not been resolved for an $\hii$ region in literature~\citep[e.g.,][]{abb+14}, the one closer to the parallax distance will be adopted \citep[see][for detail]{wba+18}. 
In addition, since the estimated kinematic distances will have large uncertainties for the sources around the Galactic Centre (GC) and Galactic anti-centre directions, we remove the $\hii$ regions located within the zones of $-$15\degr\ < {$\ell$} < 15\degr\ and 160\degr\ < {$\ell$} < 200\degr\ as suggested by \cite{wba+18}. 
Finally, we obtain kinematic distances for 184 $\hii$ regions in our sample. The line velocities and the calculated kinematic distances are listed in Table\,\ref{knownhii}.
The comparison between the parallax-based distances ($d_p$) and the calculated kinematic distances ($d_{kin}$) is shown in Fig.\,\ref{discom}.
The kinematic distances typically have larger uncertainties for the sources with $d_{\sun}<~$4\, kpc.
After an inspection of the fractional difference |$d_{kin}-d_p|/d_p$, we find that the mean and median values are 0.50 and 0.24, respectively.  These results are similar to that of \citet[][see their Table~4]{wba+18}, who has shown that the mean and median absolute fractional differences between the kinematic and parallax distances are 36\% and 26\%, respectively. In this work,
24 of the 184 $\hii$ regions have differences |$d_{kin}-d_p$| larger than 2~kpc. 
In addition, the kinematic method tends to give larger distance values for many $\hii$ regions by a mean factor of $\sim$1.5, comparable to the finding of \cite{rmz+09} and \citet{jhh+23}.
As shown in the lower panel Fig.~\ref{discom}, accurate distances of $\hii$ regions are crucial for building a reliable map of the nearby spiral arms.  

\section{Spiral structure in the solar neighbourhood}

\begin{figure*}
  \centering
  \includegraphics[width=0.95\textwidth]{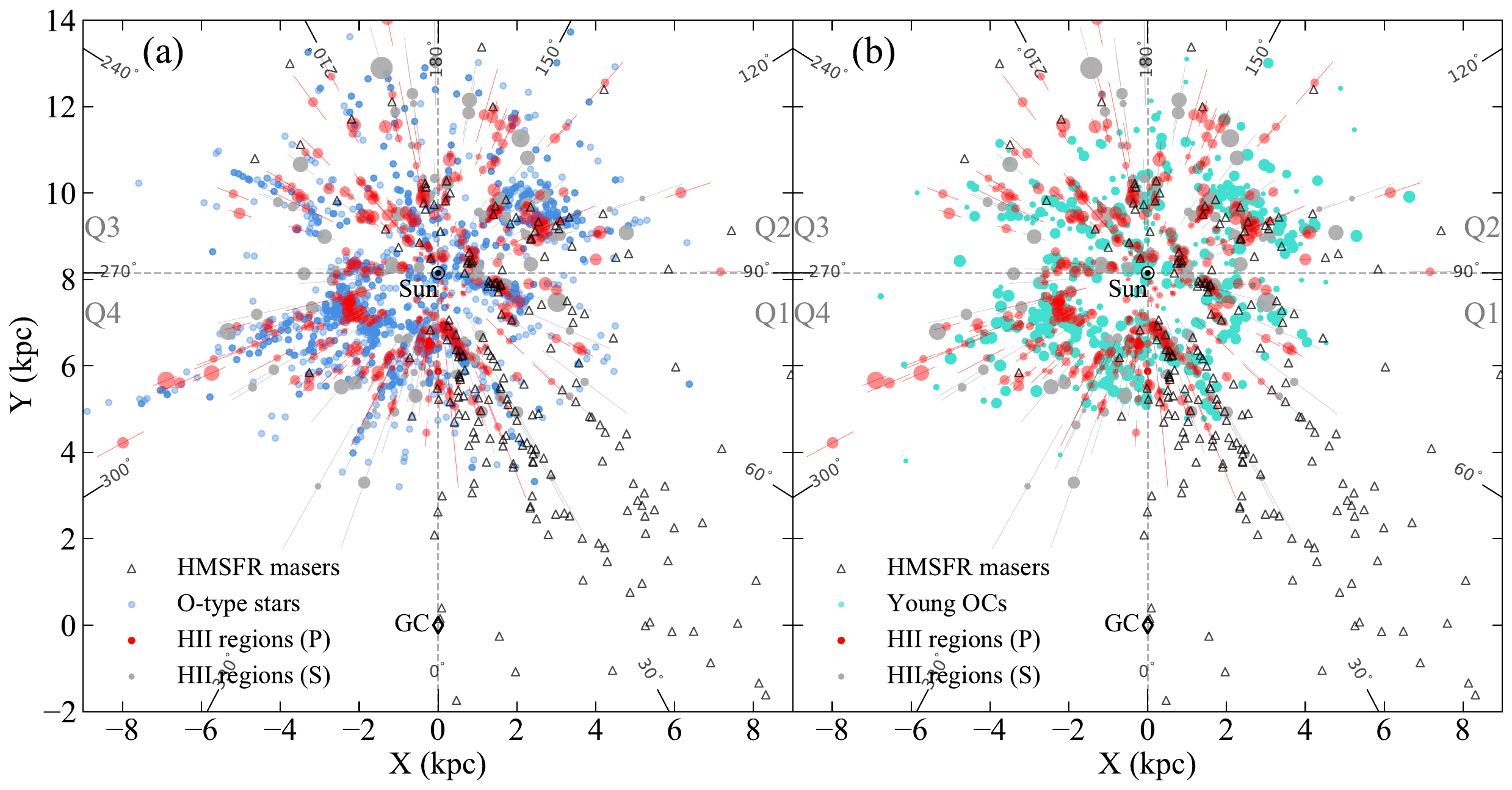}  
  \caption{Distributions of $\hii$ regions overlaid with HMSFR masers (open triangles), O-type stars (blue dots, {\it left panel}), and young OCs (cyan dots, {\it right}). An equal size of the symbol is adopted for masers and O-type stars. The dot size for each young OC is proportional to the number of member stars as \citet{hxh21}. The $\hii$ regions and the positions of the Sun and GC are the same as Fig.\,\ref{proj_hii}. }
\label{young}
\end{figure*}
\subsection{$\hii$ region distribution and comparison to other young objects}
\label{sect:sample}

To better reveal the spiral structure in the solar neighbourhood traced by $\hii$ regions, the sample size of Galactic $\hii$ regions having accurate distances should be as large as possible.
Therefore, besides the 459 $\hii$ regions given by this work, we additionally collect 12 $\hii$ regions having parallax-based distances \citep[][]{maa+22}, and the other 101 $\hii$ regions with available spectrophotometric distances \citep[][]{rus03,rag07,mdf+11,fb15}. Note that 20 of 101 $\hii$ regions have spectrophotometric distances $> 4$\,kpc, which are considered with large uncertainties as shown in Fig.\,\ref{php}, and thus should be treated with caution.
Fig.\,\ref{proj_hii} presents the projected distributions of the 572 $\hii$ regions on the Galactic plane. 
The radio luminosity of the $\hii$ regions at 5~GHz used for weighting the symbol sizes \citep[e.g.,][]{gg76,hh14} are derived according to the relation between radio luminosity and physical diameters given by \citet{pdd04}.

As shown in Fig.\,\ref{proj_hii}, the distribution of $\hii$ regions presents 
complex structures, including some cavities and overcrowded regions. 
Although the number and exact position of spiral arms are still open questions \citep[][]{fc10,xhw18,xu+23}, the overcrowded regions can be related to the segments of spiral arms referring to the commonly adopted models of Galactic spiral structure \citep[e.g., see][]{reid19,xu+23}. They are the Scutum-Centaurus Arm, the Sagittarius-Carina Arm, the Local Arm, the Perseus Arm, and the Outer Arm from the inner to the outer Galactic disc.
Besides the major spiral arm segments, some features traced by $\hii$ regions are coincident with spur or spur-like structures reported in previous works, including the spur near the direction of $\ell \sim 50^\circ$ \citep[][]{xu+16}, the Cepheus spur which starts from the Local Arm  ($X\sim1$ kpc, $Y\sim8.15$ kpc) and ends at the Perseus Arm around $\ell\sim210\degr$ \citep[][]{pmb+21}, and a spur-like structure bridging the Scutum Arm and the Sagittarius Arm \citep[][]{reid19} etc.
In addition, the structure extending from ($X\sim1.5$ kpc, $Y\sim8$ kpc) to the direction of $\ell\sim235\degr$ is well aligned with the so-called Radcliffe wave, which was first discovered as a sinusoidal vertical feature of dense molecular gas \citep{azg+20}.

\begin{figure*}
  \centering
  \includegraphics[width=0.95\textwidth]{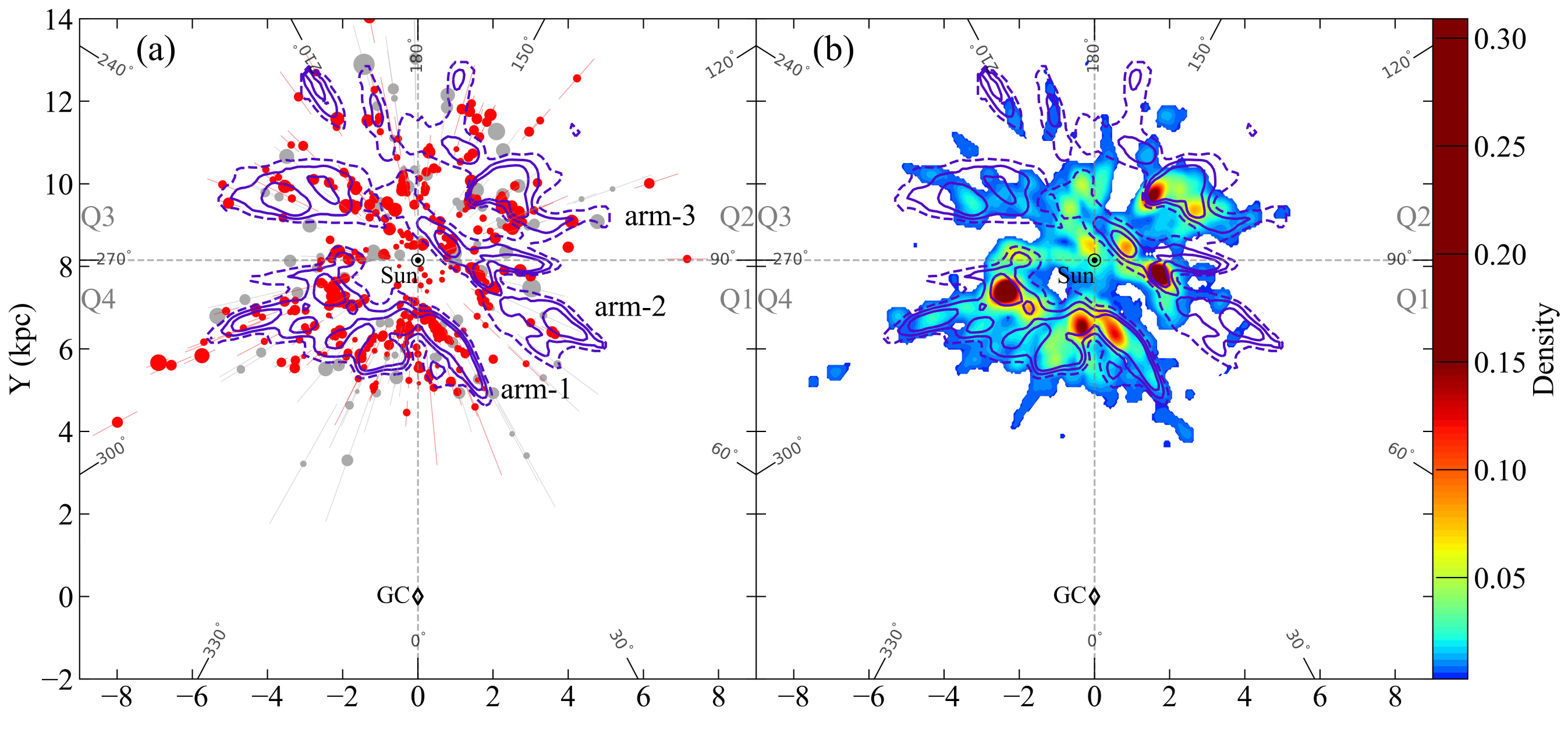} \\
   \includegraphics[width=0.95\textwidth]{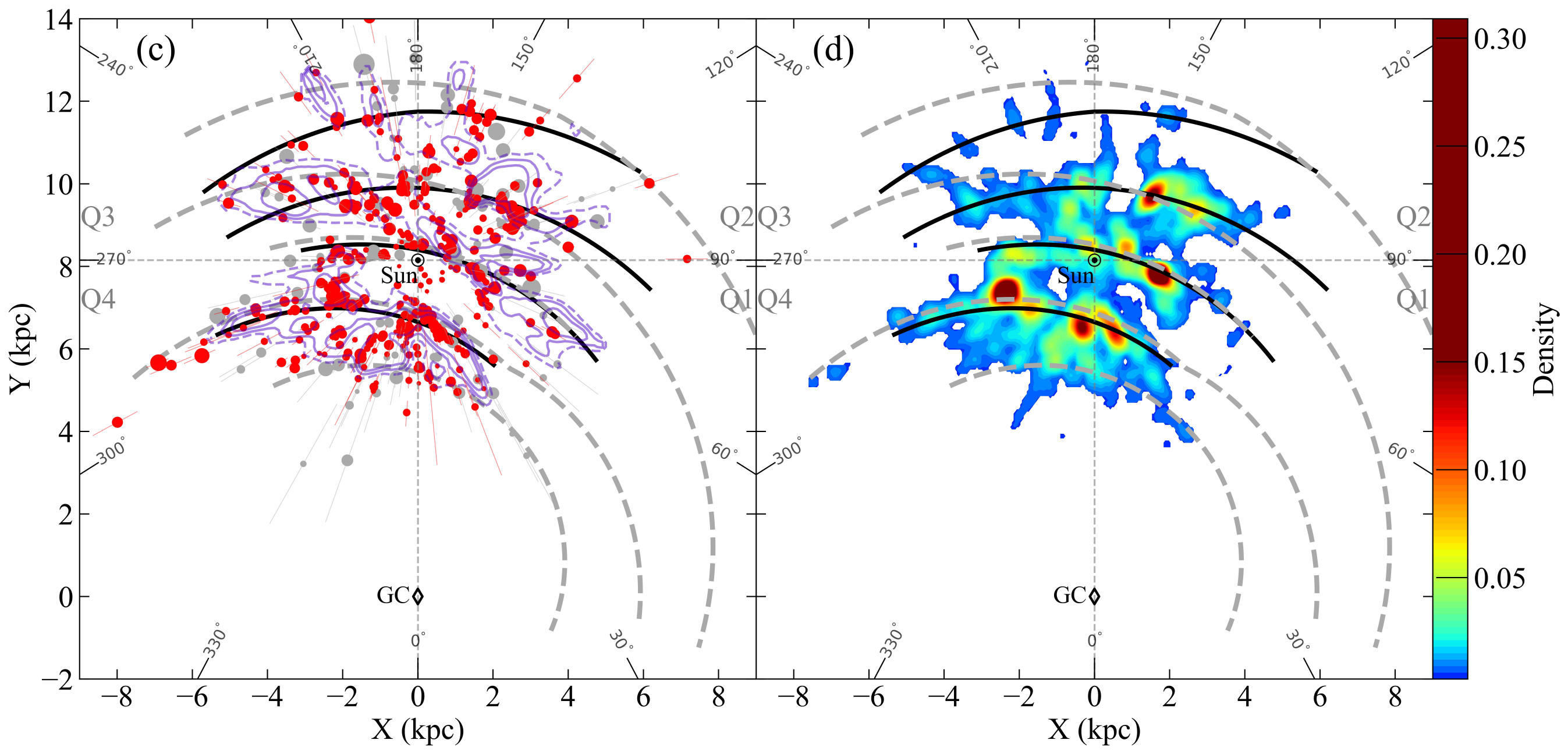} 
  \caption{{\it Top two panles}: spatial distribution of $\hii$ regions (a), and the density map of young objects (b).  The contours in panels (a), (b), and (c) represent the over-density map of upper main-sequence stars \citep{pdc+21}. The contour levels are 0 for the dashed line, as well as 0.2 and 0.4 for solid lines. Additionally, three arm-like features identified by \citep{pdc+21} are labelled in panel (a) as arm-1, arm-2, and arm-3 from the inner to the outer Galactic disc.
  {\it Bottom two panels}: similar to the upper panels, but the results are overplotted on the spiral arm models of \citet[][black solid lines]{xu+23} and \citet[][grey dashed lines]{reid19}.  The arm segments are the Outer Arm, the Perseus Arm, the Local Arm, the Carina/Sagittarius-Carina Arm, and the Scutum Arm from the outer to the inner Galactic disc.} 
\label{old}
\end{figure*}

To confirm the features shown in the distribution of $\hii$ regions, the HMSFR masers, O-type stars, and YOCs (ages\,<\,20\, Myr) with lifetimes comparable to or even younger than $\hii$ regions are collected from references to make a comparison. 
The HMSFR masers are taken from the literature discussed in Section\,3.2. In addition to the 1\,090 spectroscopy-confirmed O-type stars from \cite{xhb+21}, the extra 592 O-type stars are obtained from the updated catalogue of \cite{skiff14}. The redundancy of these O stars with those used to calculate the $\hii$ region distances has been checked and removed. The sample of 633 YOCs is derived from \cite{hxh21}. All of these YOCs have accurately determined distances based on the parallax measurements of a large number of member stars by the {\it Gaia} Early Data Release 3. The distributions of these young objects together with $\hii$ regions are given in Fig.\,\ref{young}.
As shown in the plots, the structures traced by $\hii$ regions share consistent aggregation areas indicated by HMSFR masers, O-type stars, and YOCs.
On the other hand, some cavities appearing in the $\hii$ region map are replenished by O-type stars and YOCs. For instance, the apparent gap identified in the Carina Arm ($\ell \sim 320^\circ$) is not genuine due to the limitation of the sample size of our $\hii$ regions, and this zone is enriched with some O-type stars and YOCs. 
The occurrence in the Perseus Arm ($\ell \sim 150^\circ$) may be a real cavity structure, which also appears in the results of \cite{xu+23} and \cite{gaiastruc23}.
Overall, the structures outlined by young objects with parallax-based distances do not exhibit significant deviation from each other. 
Moreover, the results also suggest that the influence of foreground/background stars for the 299 $\hii$ regions with a single candidate star (see Sect.~\ref{sect:stars}) ought to be inapparent for mapping the nearby spiral structure.

\begin{figure*}
  \centering
 \includegraphics[width=0.9\textwidth]{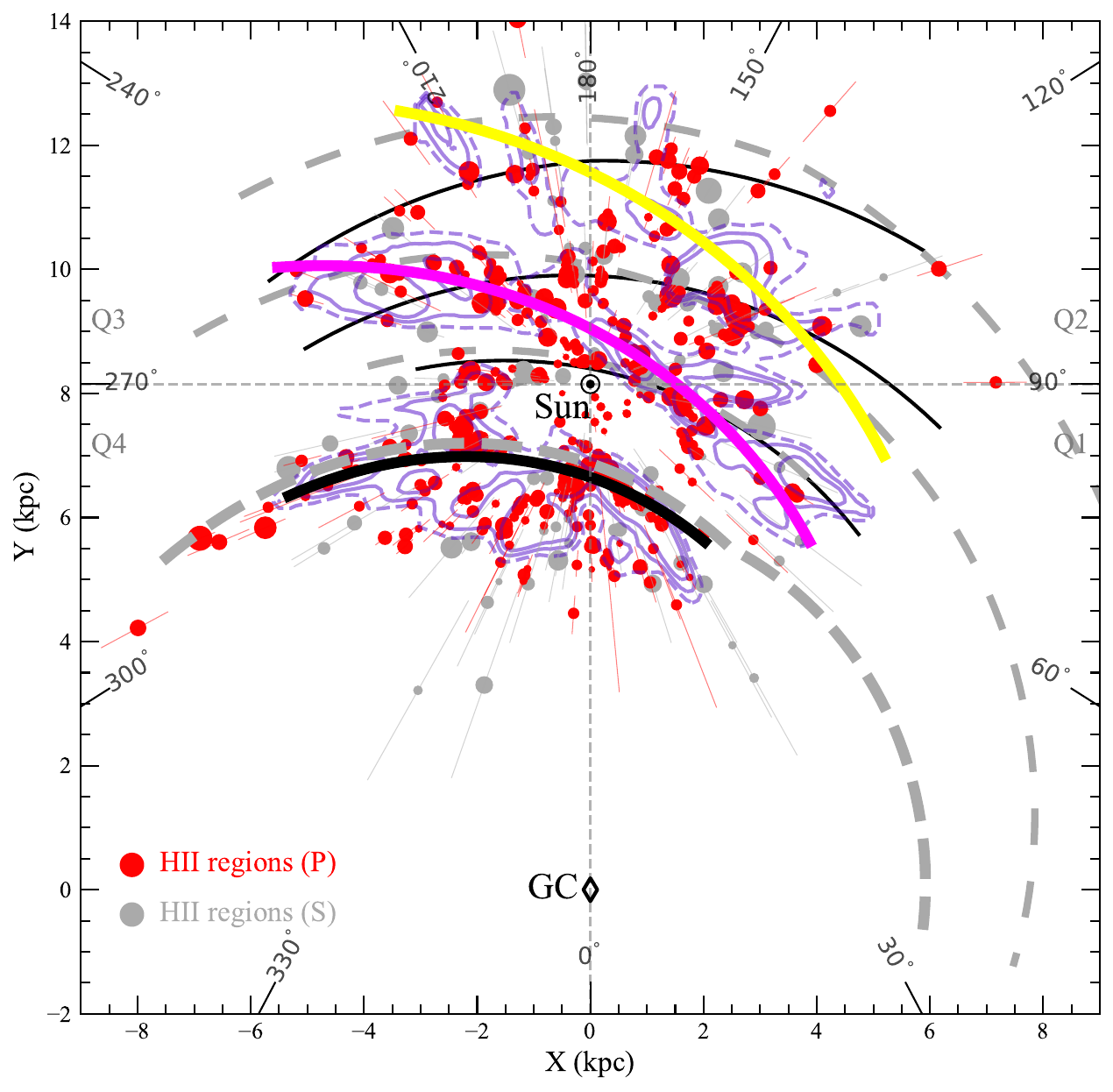} \\
  \caption{Distributions of $\hii$ regions overlaid with the corresponding spiral arm models. The fuchsia solid line traces the best-fitted Local Arm given by this work (see Section 4.2). The yellow curved line presents the Perseus Arm determined from the perturbed surface density map of $\hi$ by \cite{lev06}. The solid and long-dashed curved lines in the plot indicate the best-fitted Carina/Sagittarius-Carina Arm (thick) and the other arm segments (thin) given by \cite{xu+23} and \cite{reid19}, respectively. $\hii$ regions and the positions of the Sun and GC are the same as Fig.~\ref{proj_hii}. The over-density map of evolved stars (blue) is also shown in the plot to make a comparison.}
\label{fit}
\end{figure*}

How to derive a true picture of the Galactic spiral structure from the distributions of young objects is a problem.
The commonly adopted model of the Galactic spiral structure has been challenged by observations \citep[e.g., see][]{xu+23}. 
If we do not have prior knowledge about the overall spiral structure of the Milky Way, the relation of the overcrowded regions to the nearby spiral arms is not easily discernible from Fig.~\ref{proj_hii} and Fig.~\ref{young}. 
Some factors complicate the attempts to get an uncontroversial picture. The formation of massive stars is not confined to the major spiral arms but also occurs in the arm branches, arm spurs, and feathers as observed in spiral galaxies \citep[e.g.,][]{elme80}. Additionally, the formation sites of young objects are always scattered and/or clumped in spiral arms, resulting in patchy or bifurcate appearance of arm-like features. 
In contrast, the influence of these factors is not so serious for the widely distributed evolved stars, which dominate the mass of visible matter in the Galactic disc. 
Therefore, it will be beneficial to compare the distribution of young objects with more evolved stars to reveal the nearby spiral structure better. 
%


\subsection{Comparison with more evolved stars}
\label{stellararm}
In this work, the over-density map of upper main-sequence stars provided by \cite{pdc+21} is adopted for indicating the distribution of evolved stars for comparison. As mentioned in Sect.~\ref{sec:intro}, different works for example by using e.g., turn-off stars \citep{msk+19}, upper main-sequence stars \citep{pdc+21}, Red Clump Stars \citep{lxh22}, and B3 to B9 stars \citep[][]{glh+24} have given consistent results about the over-density map of stars in the solar neighbourhood. 
In the following, we name the three arm-like features shown in the over-density map of upper main-sequence stars as arm-1, arm-2, and arm-3 from the inner to the outer Galactic disc (see Fig.~\ref{old}) since their correspondence to the major spiral arms is still inconclusive.  

Following the method of calculating the dimensionless stellar density \citep[e.g.,][]{pdc+21,lxh22}, the local density for the position ($X, Y$) of young objects ($\hii$ regions, HMSFR masers, O-type stars, and YOCs) can be expressed as:
\begin{equation}
\Sigma(X,Y)=\frac{1}{Nh^{2}}\sum_{i=1}^{N}\biggl[K\biggl(\frac{X-x_{i}}{h}\biggr)K\biggl(\frac{Y-y_{i}}{h}\biggr)\biggr],
\end{equation}
where
\begin{equation}
K\biggl(\frac{X-x_{i}}{h}\biggr)=\frac{3}{4}\biggl(1-\biggl(\frac{X-x_{i}}{h}\biggr)^{2}\biggr).    
\end{equation}
Here, $K$ is the Epanechnikov kernel function, and $h$ stands for the kernel bandwidth (taken as 0.3 kpc), $(x_{i},y_{i})$ is the position for the $i$th young object and $N$ is the total number.
The results are given in panels (a) and (b) of Fig.\,\ref{old}. 

As shown in Fig.\,\ref{old}, both the distributions of $\hii$ regions and the density map of young objects are in good agreement with the three arm-like features appearing in the over-density map of upper main-sequence stars.
The results indicate that the spiral structures traced by young objects and evolved stars are probably consistent with each other at least in the solar neighbourhood.
Based on panels (a) and (b) of Fig.\,\ref{old}, the outer part of arm-1 is well consistent with the Sagittarius-Carina Arm defined by \citet{reid19} and \citet{xu+23}, the arm-2 overlaps with their Local Arm in the first and second Galactic quadrants but appears to be an outward feature towards the direction of 
$\ell \sim 240^\circ$ in the third Galactic quadrant, the arm-3 starts from the second Galactic quadrant and spirals outwards with a large pitch angle \citep[][]{lev06}.
The cavity structure around ($X, Y$) $\sim$ (1, 10)~kpc is interpreted as inner arm regions rather than a gap in the Perseus Arm defined by previous works.
This picture conflicts with the commonly adopted models of the Galactic spiral arms. 

The spiral arm models given by previous works \citep[e.g.,][]{reid19} for the combination of various young objects can explain the features shown in Fig.~\ref{proj_hii} and Fig.~\ref{young} (also see the lower two panels of Fig.\,\ref{old}), but deviate as large as more than 1~kpc from the two arm-like structures (arm-2 and arm-3) appearing in the over-density map of evolved stars, which dominate the mass of visible matter in the Galactic disc. 
These results imply that some modifications to these commonly used models are necessary given the following considerations:

(1) The spiral structures traced by young objects and evolved stars should not present large offsets in positions in the solar neighbourhood. The possible offsets between the spiral arms traced by young objects and old stars are about $\sim100-500$~pc at the spiral arm tangents of the Milky Way~\citep[e.g.,][]{hh15,vall22}, which should be larger than that in the regions close to the co-rotation radius if the (quasi-stationary) density wave theory \citep{lin+64,lin+66,shu+16} applies to the Milky Way Galaxy \citep[][]{rob69,frd+11}.
The solar neighbourhood is just around the Galactic co-rotation radius \citep[$\sim$8.5~kpc, e.g.,][]{dias19}. 
In addition, based on the observations towards external spiral galaxies, \citet[][]{yu+18} also suggest that the possible offsets between the two different types of spiral arms are generally small, particularly in regions near the co-rotation radius of galaxies. Except for the (quasi-stationary) density wave theory, the mechanisms for the spiral formation do not predict the offset between the spiral arms traced by young objects and evolved stars \citep[e.g.,][]{db14}.

(2) In the models of \citet{hh14}, \citet{reid19} and \citet{xu+23}, the Local Arm is extended to the direction of $\ell \sim 260^\circ$ (see Fig.\,\ref{old}) as there are clumps of young objects around ($X, Y$) $\sim$ ($-$2, 8.2)~kpc. A recent study 
\citep{glh+24} have shown that they are probably transient and younger structures compared to the three arm-like features shown in Fig.~\ref{old}.
\citet{glh+24} find that these clumps are only prominent in the distributions of O-B2 stars, and gradually become invisible when B3-B5, B6-B7, B8, and B9 stars are used to create the over-density maps. In contrast, the three arm-like features are still visible.

Based on the above considerations, we fit the arm-2 segment traced by $\hii$ regions using a commonly used logarithmic spiral model, defined as:
\begin{equation}
\ln(R/R_{\mathrm{ref}})=-(\beta-\beta_{\mathrm{ref}})\tan\psi 
\end{equation}
where $R$ and $\beta$ are the Galactocentric radius and azimuth angle, respectively. Here, $\beta$ is defined as 0$\degr$ from the Galactic Centre towards the Sun and increases with Galactic rotation. $\psi$ is a pitch angle with a reference Galactocentric radius $R_{\mathrm{ref}}$ and azimuth angle $\beta_{\mathrm{ref}}$. We obtain the best-fitted results with ($R_{\mathrm{ref}}$, $\beta_{\mathrm{ref}}$) = (9.6$\pm$
0.1 kpc, $-$7.0$\degr\pm 1.4\degr$) and $\psi$ = 25.2$\degr \pm$2.0$\degr$, as shown in Fig.~\ref{fit}. The line tracing the centre of the arm-like feature spirals outwards towards the direction of $\ell$ $\sim$240$\degr$ over 6 kpc in length in the third quadrant. 
The pitch angle of arm-2 is larger than that of the Local Arm given by previous works \citep[e.g., $\sim9^\circ-16^\circ$,][]{reid19,msk+19,hxh21,lxh22,xu+23}, but match the distributions of $\hii$ regions, other young objects and the over-density map of evolved stars.
Although the sample size is small for the third Galactic quadrant region, the arm-3 outlined by distant $\hii$ regions generally follows the Perseus Arm given by \cite{lev06} based on the distribution of $\hi$ gas, which has a pitch angle of 24$^\circ$ and is directly adopted in this work. Since arm-1 shows excellent consistency in both young objects and evolved stars, the Sagittarius-Carina Arms modelled by \cite{reid19} and \cite{xu+23} are directly adopted and shown in the plot. 

\section{Discussions}

\subsection{Evolved understanding to the Local Arm}

In this work, it is shown that the Local Arm traced by young objects is consistent with the over-density distribution of evolved stars, and probably spirals outwards towards the direction of $\ell \sim 240^\circ$ in the third Galactic quadrant, rather than towards $\ell \sim 260^\circ$. This view is not consistent with the commonly adopted spiral arm models but has been proposed in some references. 

Early studies \citep{mwc53,bf63,beck64,cggm70} have indicated the presence of three nearby spiral arms (i.e., see their Fig.1), which were suggested to be the Perseus Arm, the Orion Arm (also named as the Local Arm, the Orion-Cygnus Arm), and the Sagittarius Arm. 
The Local Arm was taken as one of the three spiral arm segments in the solar neighbourhood, having a large pitch angle of $\sim20^\circ$~\citep[e.g.,][]{aved85} and extending towards $\ell \sim 240^\circ$~\citep[e.g.,][]{mfg79}.

\citet{gg76} gave a weighted distribution of $\hii$ regions in the Galactic disc. The Orion Arm reported in previous works \citep[e.g.,][]{beck64,cggm70} became relative insignificance, which was then suggested to be a spur (the Orion spur) or a branch as seen in M\,51 or M\,101, rather than a true spiral arm. 
In the subsequent works~\citep[e.g.,][]{dwbw80,ch87,ds01,rus03,ne2001}, the Local Arm was generally taken as a secondary spiral feature, due to the less significance of the star-forming region density than that of other major arms.
During this period, different interpretations occasionally appeared in some works. For instance, \citet{vmc+08} has pointed out that the Local Arm is traced by CO and young stars towards $\ell = 240^\circ$ and extends for over 8 kpc along the line of sight. 

In the 2010s, the nature of the Local Arm has come to the fore again.
Based on the accurate trigonometric parallax method, \citet{xlr+13} analysed 30 star-forming regions in the Local Arm. The active star formation sites with an overall length of $>$5 kpc and a shallow pitch angle ($\sim10^\circ$) suggest that it is more like the adjacent Perseus and Sagittarius Arms, and extends towards $\ell \sim 260^\circ$.
The Local Arm is not likely a spur and instead may be a branch of the Perseus Arm. The results were confirmed by \citet{xu+16}, in which new observations reveal that the Local Arm is larger than previously thought, and both its pitch angle and star formation rate are comparable to those of the Galactic major spiral arms. 
The Local Arm was then extended to the fourth Galactic quadrants according to O stars with parallax measurements of Gaia DR2 by \citet{xu+2018} and of Gaia DR3 by \citet{xhb+21}. Its pitch angle is about 12$^\circ$. The results of \citet{xhb+21} were confirmed by \citet{hxh21} with YOCs and taken as the generally accepted picture of the Local Arm traced by young objects~\citep[e.g.,][]{reid19,xu+23}.

In the past few years, things got complicated when the arm-like structure (arm-2 as discussed in Sect.~\ref{stellararm}) was found in the over-density distribution of evolved stars \citep[e.g.,][]{msk+19,pdc+21,lxh22}. This structure seems to spiral outwards towards the direction of $\ell \sim 240^\circ$ in the third Galactic quadrant, rather than $\ell \sim 260^\circ$. As shown in Fig.~\ref{old}, this structure traced by evolved stars is consistent with the distributions of young objects, which motivates us to re-interpret the Local Arm referring to the early efforts~\citep[e.g.,][]{mwc53,bf63,beck64,vmc+08}.

\subsection{Radcliffe wave and the Local Arm}

The Radcliffe wave is recently discovered by \citet[][]{azg+20} as a narrow and coherent $\sim$2.7-kpc arrangement of dense gas \citep[][]{azg+20,zfl+24}, Young Stellar Objects \citep[][]{lc22}, young open clusters \citep[][]{bb24}, as well as interstellar dust, and masers, etc \citep[][]{bb22}. It is oscillating around the midplane of the Milky Way disc \citep[][]{lc22,tdp+2022}, while drifting radially away from the Galactic Centre \citep{kgz+24}. The formation mechanisms for the Radcliffe Wave remain unclear yet.

Based on the commonly used model of Galactic spiral structure \citep[][]{reid19},
\citet{azg+20} suggested that the Radcliffe wave makes up 20$\%$ of the Local Arm's width and 40$\%$ of its length, respectively, and the Local Arm crosses it at an angle of $\sim$25$\degr$. 
\cite{sad+2022} suggested that the Radcliffe wave is contained within the Local Arm proposed by \citet[][]{reid19} when considering the arm width, and it serves as a gas reservoir for the Local Arm.
\cite{bbm22} confirmed that part of the Radcliffe wave is associated with the Local Arm through an analysis of the spatial distribution of masers, radio stars, and T Tauri Stars. 
\cite{pmb+21} supposed that the Cepheus spur may be related to the Radcliffe wave. 
As shown in Fig.~\ref{old}, the location of the Cepheus spur is more consistent with a structure traced by $\hii$ regions starting from ($X, Y$) $\sim$ (1.0, 8.5)~kpc and extending towards $\ell \sim 210^\circ$, rather than the Radcliffe wave marked by \citet[][]{azg+20}.

However, if the Local Arm extends to the direction of $\ell \sim 240^\circ$, rather than towards $\ell \sim 260^\circ$, the Radcliffe wave will be in good agreement with the inner edge of this arm. If this possibility is confirmed, it will provide great potential to unveil the formation mechanisms of the Radcliffe Wave and discover more wave-like structures with more survey data leveraged.

\section{Conclusions}
In this contribution, we compiled a catalogue of $\hii$ regions with {\it Gaia} DR3 parallax-based distances and studied the nearby spiral structure using this sample as well as other young objects and evolved stars. The main results are as follows:

(1) The parallax-based distances of 459  WISE $\hii$ regions were calculated by identifying their exciting stars and adopting {\it Gaia} DR3 photogeometric/geometric distances. This catalogue presents the largest sample of $\hii$ regions with {\it Gaia} parallax-based distances in the solar neighbourhood.
These distances show excellent consistency with other parallax-based measurements and display obvious corrections compared to the spectrophotometric and kinematic techniques for many $\hii$ regions. 

(2) The distribution of $\hii$ regions exhibits complex structures (i.e., cavities and crowded regions) associated with commonly adopted spiral arm models and spur or spur-like features. We compared the distributions with other young objects (i.e., O-type stars, young open clusters, and HMSFR masers) and found that they are in good agreement with each other.
  
(3) We mapped the density in the solar neighbourhood by combining the $\hii$ region data with other young objects. Both the resulting map and distributions of $\hii$ regions show obvious correspondence with the over-density map of upper main-sequence stars.

(4) Considering the less significant offsets that exist between the spiral arms traced by young objects and evolved stars, along with the likely transient feature of clumps around {\it (X, Y)} $\sim$ ($-$2, 8.2) kpc, we proposed a new view for the nearby spiral structures. We suggested that the Local Arm likely spirals outwards towards the third Galactic quadrant with a pitch angle $\sim$ 25.2$\degr$. On the other hand, the Perseus Arm and Sagittarius-Carina Arm would align well with the spiral arm model of \cite{lev06} and the commonly adopted models of young objects of \citet{reid19} and \citet{xu+23}, respectively.

As the nearest spiral arm, the Local Arm defined by a series of studies shows a more extended feature than previously believed but remains inconclusive regarding its pitch angle. Our results indicate a Local Arm with a large pitch angle, consistent with both young objects and evolved stars. It prompts us to re-interpret the Local Arm also referring to early studies and connect it with the recently discovered Radcliffe Wave.

\section{Data availability}
Tables \ref{knownhii} and \ref{exciting stars} are only available in electronic form at the CDS via anonymous ftp to cdsarc.u-strasbg.fr (130.79.128.5) or via http://cdsweb.u-strasbg.fr/cgi-bin/qcat?J/A+A/.
\begin{acknowledgements}
We thank the anonymous referee for the instructive comments and suggestions which helped improve the manuscript.
This work is supported by the National SKA Program of China (Grant No. 2022SKA0120103), and the National Natural Science Foundation (NNSF) of China No. 11988101, 11933011, and 12103045. 
H.-L. Liu is supported  by Yunnan Fundamental Research Project (grant No. 202301AT070118,
202401AS070121), and by Xingdian Talent Support Plan--Youth Project. 
This research has made use of the VizieR catalogue access tool, CDS,
Strasbourg, France (DOI: 10.26093/cds/vizier). The original
description of the VizieR service was published in A\&AS 143, 23
This research has made use of the SIMBAD database,
operated at CDS, Strasbourg, France.

\end{acknowledgements}

\bibliographystyle{aa}
\bibliography{ams}

\begin{thebibliography}{141}
\expandafter\ifx\csname natexlab\endcsname\relax\def\natexlab#1{#1}\fi

\bibitem[{{Alves} {et~al.}(2020){Alves}, {Zucker}, {Goodman}, {Speagle}, {Meingast}, {Robitaille}, {Finkbeiner}, {Schlafly}, \& {Green}}]{azg+20}
{Alves}, J., {Zucker}, C., {Goodman}, A.~A., {et~al.} 2020, \nat, 578, 237

\bibitem[{{Alves} {et~al.}(2015){Alves}, {Calabretta}, {Davies}, {Dickinson}, {Staveley-Smith}, {Davis}, {Chen}, \& {Barr}}]{hipass}
{Alves}, M. I.~R., {Calabretta}, M., {Davies}, R.~D., {et~al.} 2015, \mnras, 450, 2025

\bibitem[{{Anderson} {et~al.}(2015){Anderson}, {Armentrout}, {Johnstone}, {Bania}, {Balser}, {Wenger}, \& {Cunningham}}]{aaj+15}
{Anderson}, L.~D., {Armentrout}, W.~P., {Johnstone}, B.~M., {et~al.} 2015, \apjs, 221, 26

\bibitem[{{Anderson} {et~al.}(2014){Anderson}, {Bania}, {Balser}, {Cunningham}, {Wenger}, {Johnstone}, \& {Armentrout}}]{abb+14}
{Anderson}, L.~D., {Bania}, T.~M., {Balser}, D.~S., {et~al.} 2014, \apjs, 212, 1

\bibitem[{{Anderson} {et~al.}(2011){Anderson}, {Bania}, {Balser}, \& {Rood}}]{abb+11}
{Anderson}, L.~D., {Bania}, T.~M., {Balser}, D.~S., \& {Rood}, R.~T. 2011, \apjs, 194, 32

\bibitem[{{Anderson} {et~al.}(2021){Anderson}, {Luisi}, {Liu}, {Wenger}, {Balser}, {Bania}, {Haffner}, {Linville}, \& {Mascoop}}]{gdigs}
{Anderson}, L.~D., {Luisi}, M., {Liu}, B., {et~al.} 2021, \apjs, 254, 28

\bibitem[{{Ard{\`e}vol} {et~al.}(2023){Ard{\`e}vol}, {Mongui{\'o}}, {Figueras}, {Romero-G{\'o}mez}, \& {Carrasco}}]{amf+23}
{Ard{\`e}vol}, J., {Mongui{\'o}}, M., {Figueras}, F., {Romero-G{\'o}mez}, M., \& {Carrasco}, J.~M. 2023, \aap, 678, A111

\bibitem[{{Armentrout} {et~al.}(2021){Armentrout}, {Anderson}, {Wenger}, {Balser}, \& {Bania}}]{aaw+21}
{Armentrout}, W.~P., {Anderson}, L.~D., {Wenger}, T.~V., {Balser}, D.~S., \& {Bania}, T.~M. 2021, \apjs, 253, 23

\bibitem[{{Astraatmadja} \& {Bailer-Jones}(2016{\natexlab{a}})}]{ab16}
{Astraatmadja}, T.~L. \& {Bailer-Jones}, C. A.~L. 2016{\natexlab{a}}, \apj, 832, 137

\bibitem[{{Astraatmadja} \& {Bailer-Jones}(2016{\natexlab{b}})}]{ab16b}
{Astraatmadja}, T.~L. \& {Bailer-Jones}, C. A.~L. 2016{\natexlab{b}}, \apj, 833, 119

\bibitem[{{Avedisova}(1985)}]{aved85}
{Avedisova}, V.~S. 1985, Soviet Astronomy Letters, 11, 185

\bibitem[{{Bailer-Jones}(2015)}]{bai15}
{Bailer-Jones}, C. A.~L. 2015, \pasp, 127, 994

\bibitem[{{Bailer-Jones} {et~al.}(2021){Bailer-Jones}, {Rybizki}, {Fouesneau}, {Demleitner}, \& {Andrae}}]{brf+21}
{Bailer-Jones}, C.~A.~L., {Rybizki}, J., {Fouesneau}, M., {Demleitner}, M., \& {Andrae}, R. 2021, \aj, 161, 147

\bibitem[{{Bailer-Jones} {et~al.}(2018){Bailer-Jones}, {Rybizki}, {Fouesneau}, {Mantelet}, \& {Andrae}}]{brf+18}
{Bailer-Jones}, C.~A.~L., {Rybizki}, J., {Fouesneau}, M., {Mantelet}, G., \& {Andrae}, R. 2018, \aj, 156, 58

\bibitem[{{Becker}(1964)}]{beck64}
{Becker}, W. 1964, \zap, 58, 202

\bibitem[{{Becker} \& {Fenkart}(1963)}]{bf63}
{Becker}, W. \& {Fenkart}, R. 1963, \zap, 56, 257

\bibitem[{{Benjamin}(2008)}]{benjamin08}
{Benjamin}, R.~A. 2008, in Astronomical Society of the Pacific Conference Series, Vol. 387, Massive Star Formation: Observations Confront Theory, ed. H.~{Beuther}, H.~{Linz}, \& T.~{Henning}, 375

\bibitem[{Bian {et~al.}(2024)Bian, Wu, Xu, Reid, Li, Zhang, Menten, Moscadelli, \& Brunthaler}]{bian+24}
Bian, S.~B., Wu, Y.~W., Xu, Y., {et~al.} 2024, The Astronomical Journal, 167, 267

\bibitem[{{Blitz} {et~al.}(1982){Blitz}, {Fich}, \& {Stark}}]{bfs82}
{Blitz}, L., {Fich}, M., \& {Stark}, A.~A. 1982, \apjs, 49, 183

\bibitem[{{Bobylev} \& {Bajkova}(2024)}]{bb24}
{Bobylev}, V.~V. \& {Bajkova}, A.~T. 2024, Research in Astronomy and Astrophysics, 24, 035010

\bibitem[{{Bobylev} {et~al.}(2022{\natexlab{a}}){Bobylev}, {Bajkova}, \& {Mishurov}}]{bbm22}
{Bobylev}, V.~V., {Bajkova}, A.~T., \& {Mishurov}, Y.~N. 2022{\natexlab{a}}, Astronomy Letters, 48, 434

\bibitem[{{Bobylev} {et~al.}(2022{\natexlab{b}}){Bobylev}, {Bajkova}, \& {Mishurov}}]{bb22}
{Bobylev}, V.~V., {Bajkova}, A.~T., \& {Mishurov}, Y.~N. 2022{\natexlab{b}}, Astrophysics, 65, 579

\bibitem[{{Bronfman} {et~al.}(1996){Bronfman}, {Nyman}, \& {May}}]{bnm96}
{Bronfman}, L., {Nyman}, L.~A., \& {May}, J. 1996, \aaps, 115, 81

\bibitem[{{Burton} \& {Gordon}(1978)}]{bg78}
{Burton}, W.~B. \& {Gordon}, M.~A. 1978, \aap, 63, 7

\bibitem[{{Cantat-Gaudin} {et~al.}(2020){Cantat-Gaudin}, {Anders}, {Castro-Ginard}, {Jordi}, {Romero-G{\'o}mez}, {Soubiran}, {Casamiquela}, {Tarricq}, {Moitinho}, {Vallenari}, {Bragaglia}, {Krone-Martins}, \& {Kounkel}}]{cac20}
{Cantat-Gaudin}, T., {Anders}, F., {Castro-Ginard}, A., {et~al.} 2020, \aap, 640, A1

\bibitem[{{Cantat-Gaudin} {et~al.}(2018){Cantat-Gaudin}, {Jordi}, {Vallenari}, {Bragaglia}, {Balaguer-N{\'u}{\~n}ez}, {Soubiran}, {Bossini}, {Moitinho}, {Castro-Ginard}, {Krone-Martins}, {Casamiquela}, {Sordo}, \& {Carrera}}]{cjv18}
{Cantat-Gaudin}, T., {Jordi}, C., {Vallenari}, A., {et~al.} 2018, \aap, 618, A93

\bibitem[{{Caswell} \& {Haynes}(1987)}]{ch87}
{Caswell}, J.~L. \& {Haynes}, R.~F. 1987, \aap, 171, 261

\bibitem[{{Chen} {et~al.}(2019){Chen}, {Huang}, {Hou}, {Tian}, {Li}, {Yuan}, {Wang}, {Wang}, {Tian}, \& {Liu}}]{chh+19}
{Chen}, B.~Q., {Huang}, Y., {Hou}, L.~G., {et~al.} 2019, \mnras, 487, 1400

\bibitem[{{Cordes} \& {Lazio}(2002)}]{ne2001}
{Cordes}, J.~M. \& {Lazio}, T.~J.~W. 2002, arXiv e-prints, astro

\bibitem[{{Court{\`e}s} {et~al.}(1970){Court{\`e}s}, {Georgelin}, {Georgelin}, \& {Monnet}}]{cggm70}
{Court{\`e}s}, G., {Georgelin}, Y.~P., {Georgelin}, Y.~M., \& {Monnet}, G. 1970, in The Spiral Structure of our Galaxy, ed. W.~{Becker} \& G.~I. {Kontopoulos}, Vol.~38, 209

\bibitem[{{Dame} {et~al.}(1987){Dame}, {Ungerechts}, {Cohen}, {de Geus}, {Grenier}, {May}, {Murphy}, {Nyman}, \& {Thaddeus}}]{dcg+87}
{Dame}, T.~M., {Ungerechts}, H., {Cohen}, R.~S., {et~al.} 1987, \apj, 322, 706

\bibitem[{{Deharveng} {et~al.}(2000){Deharveng}, {Pe{\~n}a}, {Caplan}, \& {Costero}}]{dpcc00}
{Deharveng}, L., {Pe{\~n}a}, M., {Caplan}, J., \& {Costero}, R. 2000, \mnras, 311, 329

\bibitem[{{Deharveng} {et~al.}(2010){Deharveng}, {Schuller}, {Anderson}, {Zavagno}, {Wyrowski}, {Menten}, {Bronfman}, {Testi}, {Walmsley}, \& {Wienen}}]{dsa+10}
{Deharveng}, L., {Schuller}, F., {Anderson}, L.~D., {et~al.} 2010, \aap, 523, A6

\bibitem[{{Denyshchenko} {et~al.}(2024){Denyshchenko}, {Fedorov}, {Akhmetov}, {Velichko}, \& {Dmytrenko}}]{dfa+24}
{Denyshchenko}, S.~I., {Fedorov}, P.~N., {Akhmetov}, V.~S., {Velichko}, A.~B., \& {Dmytrenko}, A.~M. 2024, \mnras, 527, 1472

\bibitem[{{Dias} {et~al.}(2019){Dias}, {Monteiro}, {L{\'e}pine}, \& {Barros}}]{dias19}
{Dias}, W.~S., {Monteiro}, H., {L{\'e}pine}, J.~R.~D., \& {Barros}, D.~A. 2019, \mnras, 486, 5726

\bibitem[{{Dias} {et~al.}(2021){Dias}, {Monteiro}, {Moitinho}, {L{\'e}pine}, {Carraro}, {Paunzen}, {Alessi}, \& {Villela}}]{dmm21}
{Dias}, W.~S., {Monteiro}, H., {Moitinho}, A., {et~al.} 2021, \mnras, 504, 356

\bibitem[{{Dobbs} \& {Baba}(2014)}]{db14}
{Dobbs}, C. \& {Baba}, J. 2014, \pasa, 31, e035

\bibitem[{{Downes} {et~al.}(1980){Downes}, {Wilson}, {Bieging}, \& {Wink}}]{dwbw80}
{Downes}, D., {Wilson}, T.~L., {Bieging}, J., \& {Wink}, J. 1980, \aaps, 40, 379

\bibitem[{{Drew} {et~al.}(2014){Drew}, {Gonzalez-Solares}, {Greimel}, {Irwin}, {K{\"u}pc{\"u} Yoldas}, {Lewis}, {Barentsen}, {Eisl{\"o}ffel}, {Farnhill}, {Martin}, {Walsh}, {Walton}, {Mohr-Smith}, {Raddi}, {Sale}, {Wright}, {Groot}, {Barlow}, {Corradi}, {Drake}, {Fabregat}, {Frew}, {G{\"a}nsicke}, {Knigge}, {Mampaso}, {Morris}, {Naylor}, {Parker}, {Phillipps}, {Ruhland}, {Steeghs}, {Unruh}, {Vink}, {Wesson}, \& {Zijlstra}}]{dgg+14}
{Drew}, J.~E., {Gonzalez-Solares}, E., {Greimel}, R., {et~al.} 2014, \mnras, 440, 2036

\bibitem[{{Drimmel} \& {Spergel}(2001)}]{ds01}
{Drimmel}, R. \& {Spergel}, D.~N. 2001, \apj, 556, 181

\bibitem[{{Elmegreen}(1980)}]{elme80}
{Elmegreen}, D.~M. 1980, \apj, 242, 528

\bibitem[{{Esteban} {et~al.}(2005){Esteban}, {Garc{\'\i}a-Rojas}, {Peimbert}, {Peimbert}, {Ruiz}, {Rodr{\'\i}guez}, \& {Carigi}}]{egp+05}
{Esteban}, C., {Garc{\'\i}a-Rojas}, J., {Peimbert}, M., {et~al.} 2005, \apjl, 618, L95

\bibitem[{{Fich} {et~al.}(1990){Fich}, {Treffers}, \& {Dahl}}]{ftd90}
{Fich}, M., {Treffers}, R.~R., \& {Dahl}, G.~P. 1990, \aj, 99, 622

\bibitem[{{Foster} \& {Brunt}(2015)}]{fb15}
{Foster}, T. \& {Brunt}, C.~M. 2015, \aj, 150, 147

\bibitem[{{Foster} \& {Cooper}(2010)}]{fc10}
{Foster}, T. \& {Cooper}, B. 2010, in Astronomical Society of the Pacific Conference Series, Vol. 438, The Dynamic Interstellar Medium: A Celebration of the Canadian Galactic Plane Survey, ed. R.~{Kothes}, T.~L. {Landecker}, \& A.~G. {Willis}, 16

\bibitem[{{Foyle} {et~al.}(2011){Foyle}, {Rix}, {Dobbs}, {Leroy}, \& {Walter}}]{frd+11}
{Foyle}, K., {Rix}, H.~W., {Dobbs}, C.~L., {Leroy}, A.~K., \& {Walter}, F. 2011, \apj, 735, 101

\bibitem[{{Gaia Collaboration} {et~al.}(2018){Gaia Collaboration}, {Brown}, {Vallenari}, {Prusti}, {de Bruijne}, {Babusiaux}, {Bailer-Jones}, {Biermann}, {Evans}, {Eyer}, {Jansen}, {Jordi}, {Klioner}, {Lammers}, {Lindegren}, {Luri}, {Mignard}, {Panem}, {Pourbaix}, {Randich}, {Sartoretti}, {Siddiqui}, {Soubiran}, {van Leeuwen}, {Walton}, {Arenou}, {Bastian}, {Cropper}, {Drimmel}, {Katz}, {Lattanzi}, {Bakker}, {Cacciari}, {Casta{\~n}eda}, {Chaoul}, {Cheek}, {De Angeli}, {Fabricius}, {Guerra}, {Holl}, {Masana}, {Messineo}, {Mowlavi}, {Nienartowicz}, {Panuzzo}, {Portell}, {Riello}, {Seabroke}, {Tanga}, {Th{\'e}venin}, {Gracia-Abril}, {Comoretto}, {Garcia-Reinaldos}, {Teyssier}, {Altmann}, {Andrae}, {Audard}, {Bellas-Velidis}, {Benson}, {Berthier}, {Blomme}, {Burgess}, {Busso}, {Carry}, {Cellino}, {Clementini}, {Clotet}, {Creevey}, {Davidson}, {De Ridder}, {Delchambre}, {Dell'Oro}, {Ducourant}, {Fern{\'a}ndez-Hern{\'a}ndez}, {Fouesneau}, {Fr{\'e}mat}, {Galluccio}, {Garc{\'\i}a-Torres},
  {Gonz{\'a}lez-N{\'u}{\~n}ez}, {Gonz{\'a}lez-Vidal}, {Gosset}, {Guy}, {Halbwachs}, {Hambly}, {Harrison}, {Hern{\'a}ndez}, {Hestroffer}, {Hodgkin}, {Hutton}, {Jasniewicz}, {Jean-Antoine-Piccolo}, {Jordan}, {Korn}, {Krone-Martins}, {Lanzafame}, {Lebzelter}, {L{\"o}ffler}, {Manteiga}, {Marrese}, {Mart{\'\i}n-Fleitas}, {Moitinho}, {Mora}, {Muinonen}, {Osinde}, {Pancino}, {Pauwels}, {Petit}, {Recio-Blanco}, {Richards}, {Rimoldini}, {Robin}, {Sarro}, {Siopis}, {Smith}, {Sozzetti}, {S{\"u}veges}, {Torra}, {van Reeven}, {Abbas}, {Abreu Aramburu}, {Accart}, {Aerts}, {Altavilla}, {{\'A}lvarez}, {Alvarez}, {Alves}, {Anderson}, {Andrei}, {Anglada Varela}, {Antiche}, {Antoja}, {Arcay}, {Astraatmadja}, {Bach}, {Baker}, {Balaguer-N{\'u}{\~n}ez}, {Balm}, {Barache}, {Barata}, {Barbato}, {Barblan}, {Barklem}, {Barrado}, {Barros}, {Barstow}, {Bartholom{\'e} Mu{\~n}oz}, {Bassilana}, {Becciani}, {Bellazzini}, {Berihuete}, {Bertone}, {Bianchi}, {Bienaym{\'e}}, {Blanco-Cuaresma}, {Boch}, {Boeche}, {Bombrun}, {Borrachero},
  {Bossini}, {Bouquillon}, {Bourda}, {Bragaglia}, {Bramante}, {Breddels}, {Bressan}, {Brouillet}, {Br{\"u}semeister}, {Brugaletta}, {Bucciarelli}, {Burlacu}, {Busonero}, {Butkevich}, {Buzzi}, {Caffau}, {Cancelliere}, {Cannizzaro}, {Cantat-Gaudin}, {Carballo}, {Carlucci}, {Carrasco}, {Casamiquela}, {Castellani}, {Castro-Ginard}, {Charlot}, {Chemin}, {Chiavassa}, {Cocozza}, {Costigan}, {Cowell}, {Crifo}, {Crosta}, {Crowley}, {Cuypers}, {Dafonte}, {Damerdji}, {Dapergolas}, {David}, {David}, {de Laverny}, {De Luise}, {De March}, {de Martino}, {de Souza}, {de Torres}, {Debosscher}, {del Pozo}, {Delbo}, {Delgado}, {Delgado}, {Di Matteo}, {Diakite}, {Diener}, {Distefano}, {Dolding}, {Drazinos}, {Dur{\'a}n}, {Edvardsson}, {Enke}, {Eriksson}, {Esquej}, {Eynard Bontemps}, {Fabre}, {Fabrizio}, {Faigler}, {Falc{\~a}o}, {Farr{\`a}s Casas}, {Federici}, {Fedorets}, {Fernique}, {Figueras}, {Filippi}, {Findeisen}, {Fonti}, {Fraile}, {Fraser}, {Fr{\'e}zouls}, {Gai}, {Galleti}, {Garabato}, {Garc{\'\i}a-Sedano}, {Garofalo},
  {Garralda}, {Gavel}, {Gavras}, {Gerssen}, {Geyer}, {Giacobbe}, {Gilmore}, {Girona}, {Giuffrida}, {Glass}, {Gomes}, {Granvik}, {Gueguen}, {Guerrier}, {Guiraud}, {Guti{\'e}rrez-S{\'a}nchez}, {Haigron}, {Hatzidimitriou}, {Hauser}, {Haywood}, {Heiter}, {Helmi}, {Heu}, {Hilger}, {Hobbs}, {Hofmann}, {Holland}, {Huckle}, {Hypki}, {Icardi}, {Jan{\ss}en}, {Jevardat de Fombelle}, {Jonker}, {Juh{\'a}sz}, {Julbe}, {Karampelas}, {Kewley}, {Klar}, {Kochoska}, {Kohley}, {Kolenberg}, {Kontizas}, {Kontizas}, {Koposov}, {Kordopatis}, {Kostrzewa-Rutkowska}, {Koubsky}, {Lambert}, {Lanza}, {Lasne}, {Lavigne}, {Le Fustec}, {Le Poncin-Lafitte}, {Lebreton}, {Leccia}, {Leclerc}, {Lecoeur-Taibi}, {Lenhardt}, {Leroux}, {Liao}, {Licata}, {Lindstr{\o}m}, {Lister}, {Livanou}, {Lobel}, {L{\'o}pez}, {Managau}, {Mann}, {Mantelet}, {Marchal}, {Marchant}, {Marconi}, {Marinoni}, {Marschalk{\'o}}, {Marshall}, {Martino}, {Marton}, {Mary}, {Massari}, {Matijevi{\v{c}}}, {Mazeh}, {McMillan}, {Messina}, {Michalik}, {Millar}, {Molina}, {Molinaro},
  {Moln{\'a}r}, {Montegriffo}, {Mor}, {Morbidelli}, {Morel}, {Morris}, {Mulone}, {Muraveva}, {Musella}, {Nelemans}, {Nicastro}, {Noval}, {O'Mullane}, {Ord{\'e}novic}, {Ord{\'o}{\~n}ez-Blanco}, {Osborne}, {Pagani}, {Pagano}, {Pailler}, {Palacin}, {Palaversa}, {Panahi}, {Pawlak}, {Piersimoni}, {Pineau}, {Plachy}, {Plum}, {Poggio}, {Poujoulet}, {Pr{\v{s}}a}, {Pulone}, {Racero}, {Ragaini}, {Rambaux}, {Ramos-Lerate}, {Regibo}, {Reyl{\'e}}, {Riclet}, {Ripepi}, {Riva}, {Rivard}, {Rixon}, {Roegiers}, {Roelens}, {Romero-G{\'o}mez}, {Rowell}, {Royer}, {Ruiz-Dern}, {Sadowski}, {Sagrist{\`a} Sell{\'e}s}, {Sahlmann}, {Salgado}, {Salguero}, {Sanna}, {Santana-Ros}, {Sarasso}, {Savietto}, {Schultheis}, {Sciacca}, {Segol}, {Segovia}, {S{\'e}gransan}, {Shih}, {Siltala}, {Silva}, {Smart}, {Smith}, {Solano}, {Solitro}, {Sordo}, {Soria Nieto}, {Souchay}, {Spagna}, {Spoto}, {Stampa}, {Steele}, {Steidelm{\"u}ller}, {Stephenson}, {Stoev}, {Suess}, {Surdej}, {Szabados}, {Szegedi-Elek}, {Tapiador}, {Taris}, {Tauran}, {Taylor},
  {Teixeira}, {Terrett}, {Teyssandier}, {Thuillot}, {Titarenko}, {Torra Clotet}, {Turon}, {Ulla}, {Utrilla}, {Uzzi}, {Vaillant}, {Valentini}, {Valette}, {van Elteren}, {Van Hemelryck}, {van Leeuwen}, {Vaschetto}, {Vecchiato}, {Veljanoski}, {Viala}, {Vicente}, {Vogt}, {von Essen}, {Voss}, {Votruba}, {Voutsinas}, {Walmsley}, {Weiler}, {Wertz}, {Wevers}, {Wyrzykowski}, {Yoldas}, {{\v{Z}}erjal}, {Ziaeepour}, {Zorec}, {Zschocke}, {Zucker}, {Zurbach}, \& {Zwitter}}]{gaia18}
{Gaia Collaboration}, {Brown}, A.~G.~A., {Vallenari}, A., {et~al.} 2018, \aap, 616, A1

\bibitem[{{Gaia Collaboration} {et~al.}(2023{\natexlab{a}}){Gaia Collaboration}, {Drimmel}, {Romero-G{\'o}mez}, {Chemin}, {Ramos}, {Poggio}, {Ripepi}, {Andrae}, {Blomme}, {Cantat-Gaudin}, {Castro-Ginard}, {Clementini}, {Figueras}, {Fouesneau}, {Fr{\'e}mat}, {Jardine}, {Khanna}, {Lobel}, {Marshall}, {Muraveva}, {Brown}, {Vallenari}, {Prusti}, {de Bruijne}, {Arenou}, {Babusiaux}, {Biermann}, {Creevey}, {Ducourant}, {Evans}, {Eyer}, {Guerra}, {Hutton}, {Jordi}, {Klioner}, {Lammers}, {Lindegren}, {Luri}, {Mignard}, {Panem}, {Pourbaix}, {Randich}, {Sartoretti}, {Soubiran}, {Tanga}, {Walton}, {Bailer-Jones}, {Bastian}, {Jansen}, {Katz}, {Lattanzi}, {van Leeuwen}, {Bakker}, {Cacciari}, {Casta{\~n}eda}, {De Angeli}, {Fabricius}, {Galluccio}, {Guerrier}, {Heiter}, {Masana}, {Messineo}, {Mowlavi}, {Nicolas}, {Nienartowicz}, {Pailler}, {Panuzzo}, {Riclet}, {Roux}, {Seabroke}, {Sordo}, {Th{\'e}venin}, {Gracia-Abril}, {Portell}, {Teyssier}, {Altmann}, {Audard}, {Bellas-Velidis}, {Benson}, {Berthier}, {Burgess},
  {Busonero}, {Busso}, {C{\'a}novas}, {Carry}, {Cellino}, {Cheek}, {Damerdji}, {Davidson}, {de Teodoro}, {Nu{\~n}ez Campos}, {Delchambre}, {Dell'Oro}, {Esquej}, {Fern{\'a}ndez-Hern{\'a}ndez}, {Fraile}, {Garabato}, {Garc{\'\i}a-Lario}, {Gosset}, {Haigron}, {Halbwachs}, {Hambly}, {Harrison}, {Hern{\'a}ndez}, {Hestroffer}, {Hodgkin}, {Holl}, {Jan{\ss}en}, {Jevardat de Fombelle}, {Jordan}, {Krone-Martins}, {Lanzafame}, {L{\"o}ffler}, {Marchal}, {Marrese}, {Moitinho}, {Muinonen}, {Osborne}, {Pancino}, {Pauwels}, {Recio-Blanco}, {Reyl{\'e}}, {Riello}, {Rimoldini}, {Roegiers}, {Rybizki}, {Sarro}, {Siopis}, {Smith}, {Sozzetti}, {Utrilla}, {van Leeuwen}, {Abbas}, {{\'A}brah{\'a}m}, {Abreu Aramburu}, {Aerts}, {Aguado}, {Ajaj}, {Aldea-Montero}, {Altavilla}, {{\'A}lvarez}, {Alves}, {Anders}, {Anderson}, {Anglada Varela}, {Antoja}, {Baines}, {Baker}, {Balaguer-N{\'u}{\~n}ez}, {Balbinot}, {Balog}, {Barache}, {Barbato}, {Barros}, {Barstow}, {Bartolom{\'e}}, {Bassilana}, {Bauchet}, {Becciani}, {Bellazzini}, {Berihuete},
  {Bernet}, {Bertone}, {Bianchi}, {Binnenfeld}, {Blanco-Cuaresma}, {Boch}, {Bombrun}, {Bossini}, {Bouquillon}, {Bragaglia}, {Bramante}, {Breedt}, {Bressan}, {Brouillet}, {Brugaletta}, {Bucciarelli}, {Burlacu}, {Butkevich}, {Buzzi}, {Caffau}, {Cancelliere}, {Carballo}, {Carlucci}, {Carnerero}, {Carrasco}, {Casamiquela}, {Castellani}, {Chaoul}, {Charlot}, {Chiaramida}, {Chiavassa}, {Chornay}, {Comoretto}, {Contursi}, {Cooper}, {Cornez}, {Cowell}, {Crifo}, {Cropper}, {Crosta}, {Crowley}, {Dafonte}, {Dapergolas}, {David}, {de Laverny}, {De Luise}, {De March}, {De Ridder}, {de Souza}, {de Torres}, {del Peloso}, {del Pozo}, {Delbo}, {Delgado}, {Delisle}, {Demouchy}, {Dharmawardena}, {Di Matteo}, {Diakite}, {Diener}, {Distefano}, {Dolding}, {Enke}, {Fabre}, {Fabrizio}, {Faigler}, {Fedorets}, {Fernique}, {Fournier}, {Fouron}, {Fragkoudi}, {Gai}, {Garcia-Gutierrez}, {Garcia-Reinaldos}, {Garc{\'\i}a-Torres}, {Garofalo}, {Gavel}, {Gavras}, {Gerlach}, {Geyer}, {Giacobbe}, {Gilmore}, {Girona}, {Giuffrida}, {Gomel},
  {Gomez}, {Gonz{\'a}lez-N{\'u}{\~n}ez}, {Gonz{\'a}lez-Santamar{\'\i}a}, {Gonz{\'a}lez-Vidal}, {Granvik}, {Guillout}, {Guiraud}, {Guti{\'e}rrez-S{\'a}nchez}, {Guy}, {Hatzidimitriou}, {Hauser}, {Haywood}, {Helmer}, {Helmi}, {Sarmiento}, {Hidalgo}, {H{\l}adczuk}, {Hobbs}, {Holland}, {Huckle}, {Jasniewicz}, {Jean-Antoine Piccolo}, {Jim{\'e}nez-Arranz}, {Juaristi Campillo}, {Julbe}, {Karbevska}, {Kervella}, {Kordopatis}, {Korn}, {K{\'o}sp{\'a}l}, {Kostrzewa-Rutkowska}, {Kruszy{\'n}ska}, {Kun}, {Laizeau}, {Lambert}, {Lanza}, {Lasne}, {Le Campion}, {Lebreton}, {Lebzelter}, {Leccia}, {Leclerc}, {Lecoeur-Taibi}, {Liao}, {Licata}, {Lindstr{\o}m}, {Lister}, {Livanou}, {Lorca}, {Loup}, {Madrero Pardo}, {Magdaleno Romeo}, {Managau}, {Mann}, {Manteiga}, {Marchant}, {Marconi}, {Marcos}, {Marcos Santos}, {Mar{\'\i}n Pina}, {Marinoni}, {Marocco}, {Martin Polo}, {Mart{\'\i}n-Fleitas}, {Marton}, {Mary}, {Masip}, {Massari}, {Mastrobuono-Battisti}, {Mazeh}, {McMillan}, {Messina}, {Michalik}, {Millar}, {Mints}, {Molina},
  {Molinaro}, {Moln{\'a}r}, {Monari}, {Mongui{\'o}}, {Montegriffo}, {Montero}, {Mor}, {Mora}, {Morbidelli}, {Morel}, {Morris}, {Murphy}, {Musella}, {Nagy}, {Noval}, {Oca{\~n}a}, {Ogden}, {Ordenovic}, {Osinde}, {Pagani}, {Pagano}, {Palaversa}, {Palicio}, {Pallas-Quintela}, {Panahi}, {Payne-Wardenaar}, {Pe{\~n}alosa Esteller}, {Penttil{\"a}}, {Pichon}, {Piersimoni}, {Pineau}, {Plachy}, {Plum}, {Pr{\v{s}}a}, {Pulone}, {Racero}, {Ragaini}, {Rainer}, {Raiteri}, {Ramos-Lerate}, {Re Fiorentin}, {Regibo}, {Richards}, {Rios Diaz}, {Riva}, {Rix}, {Rixon}, {Robichon}, {Robin}, {Robin}, {Roelens}, {Rogues}, {Rohrbasser}, {Rowell}, {Royer}, {Ruz Mieres}, {Rybicki}, {Sadowski}, {S{\'a}ez N{\'u}{\~n}ez}, {Sagrist{\`a} Sell{\'e}s}, {Sahlmann}, {Salguero}, {Samaras}, {Sanchez Gimenez}, {Sanna}, {Santove{\~n}a}, {Sarasso}, {Schultheis}, {Sciacca}, {Segol}, {Segovia}, {S{\'e}gransan}, {Semeux}, {Shahaf}, {Siddiqui}, {Siebert}, {Siltala}, {Silvelo}, {Slezak}, {Slezak}, {Smart}, {Snaith}, {Solano}, {Solitro}, {Souami}, {Souchay},
  {Spagna}, {Spina}, {Spoto}, {Steele}, {Steidelm{\"u}ller}, {Stephenson}, {S{\"u}veges}, {Surdej}, {Szabados}, {Szegedi-Elek}, {Taris}, {Taylor}, {Teixeira}, {Tolomei}, {Tonello}, {Torra}, {Torra}, {Torralba Elipe}, {Trabucchi}, {Tsounis}, {Turon}, {Ulla}, {Unger}, {Vaillant}, {van Dillen}, {van Reeven}, {Vanel}, {Vecchiato}, {Viala}, {Vicente}, {Voutsinas}, {Weiler}, {Wevers}, {Wyrzykowski}, {Yoldas}, {Yvard}, {Zhao}, {Zorec}, {Zucker}, \& {Zwitter}}]{gaiastruc23}
{Gaia Collaboration}, {Drimmel}, R., {Romero-G{\'o}mez}, M., {et~al.} 2023{\natexlab{a}}, \aap, 674, A37

\bibitem[{{Gaia Collaboration} {et~al.}(2023{\natexlab{b}}){Gaia Collaboration}, {Vallenari}, {Brown}, {Prusti}, {de Bruijne}, {Arenou}, {Babusiaux}, {Biermann}, {Creevey}, {Ducourant}, {Evans}, {Eyer}, {Guerra}, {Hutton}, {Jordi}, {Klioner}, {Lammers}, {Lindegren}, {Luri}, {Mignard}, {Panem}, {Pourbaix}, {Randich}, {Sartoretti}, {Soubiran}, {Tanga}, {Walton}, {Bailer-Jones}, {Bastian}, {Drimmel}, {Jansen}, {Katz}, {Lattanzi}, {van Leeuwen}, {Bakker}, {Cacciari}, {Casta{\~n}eda}, {De Angeli}, {Fabricius}, {Fouesneau}, {Fr{\'e}mat}, {Galluccio}, {Guerrier}, {Heiter}, {Masana}, {Messineo}, {Mowlavi}, {Nicolas}, {Nienartowicz}, {Pailler}, {Panuzzo}, {Riclet}, {Roux}, {Seabroke}, {Sordo}, {Th{\'e}venin}, {Gracia-Abril}, {Portell}, {Teyssier}, {Altmann}, {Andrae}, {Audard}, {Bellas-Velidis}, {Benson}, {Berthier}, {Blomme}, {Burgess}, {Busonero}, {Busso}, {C{\'a}novas}, {Carry}, {Cellino}, {Cheek}, {Clementini}, {Damerdji}, {Davidson}, {de Teodoro}, {Nu{\~n}ez Campos}, {Delchambre}, {Dell'Oro}, {Esquej},
  {Fern{\'a}ndez-Hern{\'a}ndez}, {Fraile}, {Garabato}, {Garc{\'\i}a-Lario}, {Gosset}, {Haigron}, {Halbwachs}, {Hambly}, {Harrison}, {Hern{\'a}ndez}, {Hestroffer}, {Hodgkin}, {Holl}, {Jan{\ss}en}, {Jevardat de Fombelle}, {Jordan}, {Krone-Martins}, {Lanzafame}, {L{\"o}ffler}, {Marchal}, {Marrese}, {Moitinho}, {Muinonen}, {Osborne}, {Pancino}, {Pauwels}, {Recio-Blanco}, {Reyl{\'e}}, {Riello}, {Rimoldini}, {Roegiers}, {Rybizki}, {Sarro}, {Siopis}, {Smith}, {Sozzetti}, {Utrilla}, {van Leeuwen}, {Abbas}, {{\'A}brah{\'a}m}, {Abreu Aramburu}, {Aerts}, {Aguado}, {Ajaj}, {Aldea-Montero}, {Altavilla}, {{\'A}lvarez}, {Alves}, {Anders}, {Anderson}, {Anglada Varela}, {Antoja}, {Baines}, {Baker}, {Balaguer-N{\'u}{\~n}ez}, {Balbinot}, {Balog}, {Barache}, {Barbato}, {Barros}, {Barstow}, {Bartolom{\'e}}, {Bassilana}, {Bauchet}, {Becciani}, {Bellazzini}, {Berihuete}, {Bernet}, {Bertone}, {Bianchi}, {Binnenfeld}, {Blanco-Cuaresma}, {Blazere}, {Boch}, {Bombrun}, {Bossini}, {Bouquillon}, {Bragaglia}, {Bramante}, {Breedt},
  {Bressan}, {Brouillet}, {Brugaletta}, {Bucciarelli}, {Burlacu}, {Butkevich}, {Buzzi}, {Caffau}, {Cancelliere}, {Cantat-Gaudin}, {Carballo}, {Carlucci}, {Carnerero}, {Carrasco}, {Casamiquela}, {Castellani}, {Castro-Ginard}, {Chaoul}, {Charlot}, {Chemin}, {Chiaramida}, {Chiavassa}, {Chornay}, {Comoretto}, {Contursi}, {Cooper}, {Cornez}, {Cowell}, {Crifo}, {Cropper}, {Crosta}, {Crowley}, {Dafonte}, {Dapergolas}, {David}, {David}, {de Laverny}, {De Luise}, {De March}, {De Ridder}, {de Souza}, {de Torres}, {del Peloso}, {del Pozo}, {Delbo}, {Delgado}, {Delisle}, {Demouchy}, {Dharmawardena}, {Di Matteo}, {Diakite}, {Diener}, {Distefano}, {Dolding}, {Edvardsson}, {Enke}, {Fabre}, {Fabrizio}, {Faigler}, {Fedorets}, {Fernique}, {Fienga}, {Figueras}, {Fournier}, {Fouron}, {Fragkoudi}, {Gai}, {Garcia-Gutierrez}, {Garcia-Reinaldos}, {Garc{\'\i}a-Torres}, {Garofalo}, {Gavel}, {Gavras}, {Gerlach}, {Geyer}, {Giacobbe}, {Gilmore}, {Girona}, {Giuffrida}, {Gomel}, {Gomez}, {Gonz{\'a}lez-N{\'u}{\~n}ez},
  {Gonz{\'a}lez-Santamar{\'\i}a}, {Gonz{\'a}lez-Vidal}, {Granvik}, {Guillout}, {Guiraud}, {Guti{\'e}rrez-S{\'a}nchez}, {Guy}, {Hatzidimitriou}, {Hauser}, {Haywood}, {Helmer}, {Helmi}, {Sarmiento}, {Hidalgo}, {Hilger}, {H{\l}adczuk}, {Hobbs}, {Holland}, {Huckle}, {Jardine}, {Jasniewicz}, {Jean-Antoine Piccolo}, {Jim{\'e}nez-Arranz}, {Jorissen}, {Juaristi Campillo}, {Julbe}, {Karbevska}, {Kervella}, {Khanna}, {Kontizas}, {Kordopatis}, {Korn}, {K{\'o}sp{\'a}l}, {Kostrzewa-Rutkowska}, {Kruszy{\'n}ska}, {Kun}, {Laizeau}, {Lambert}, {Lanza}, {Lasne}, {Le Campion}, {Lebreton}, {Lebzelter}, {Leccia}, {Leclerc}, {Lecoeur-Taibi}, {Liao}, {Licata}, {Lindstr{\o}m}, {Lister}, {Livanou}, {Lobel}, {Lorca}, {Loup}, {Madrero Pardo}, {Magdaleno Romeo}, {Managau}, {Mann}, {Manteiga}, {Marchant}, {Marconi}, {Marcos}, {Marcos Santos}, {Mar{\'\i}n Pina}, {Marinoni}, {Marocco}, {Marshall}, {Martin Polo}, {Mart{\'\i}n-Fleitas}, {Marton}, {Mary}, {Masip}, {Massari}, {Mastrobuono-Battisti}, {Mazeh}, {McMillan}, {Messina}, {Michalik},
  {Millar}, {Mints}, {Molina}, {Molinaro}, {Moln{\'a}r}, {Monari}, {Mongui{\'o}}, {Montegriffo}, {Montero}, {Mor}, {Mora}, {Morbidelli}, {Morel}, {Morris}, {Muraveva}, {Murphy}, {Musella}, {Nagy}, {Noval}, {Oca{\~n}a}, {Ogden}, {Ordenovic}, {Osinde}, {Pagani}, {Pagano}, {Palaversa}, {Palicio}, {Pallas-Quintela}, {Panahi}, {Payne-Wardenaar}, {Pe{\~n}alosa Esteller}, {Penttil{\"a}}, {Pichon}, {Piersimoni}, {Pineau}, {Plachy}, {Plum}, {Poggio}, {Pr{\v{s}}a}, {Pulone}, {Racero}, {Ragaini}, {Rainer}, {Raiteri}, {Rambaux}, {Ramos}, {Ramos-Lerate}, {Re Fiorentin}, {Regibo}, {Richards}, {Rios Diaz}, {Ripepi}, {Riva}, {Rix}, {Rixon}, {Robichon}, {Robin}, {Robin}, {Roelens}, {Rogues}, {Rohrbasser}, {Romero-G{\'o}mez}, {Rowell}, {Royer}, {Ruz Mieres}, {Rybicki}, {Sadowski}, {S{\'a}ez N{\'u}{\~n}ez}, {Sagrist{\`a} Sell{\'e}s}, {Sahlmann}, {Salguero}, {Samaras}, {Sanchez Gimenez}, {Sanna}, {Santove{\~n}a}, {Sarasso}, {Schultheis}, {Sciacca}, {Segol}, {Segovia}, {S{\'e}gransan}, {Semeux}, {Shahaf}, {Siddiqui}, {Siebert},
  {Siltala}, {Silvelo}, {Slezak}, {Slezak}, {Smart}, {Snaith}, {Solano}, {Solitro}, {Souami}, {Souchay}, {Spagna}, {Spina}, {Spoto}, {Steele}, {Steidelm{\"u}ller}, {Stephenson}, {S{\"u}veges}, {Surdej}, {Szabados}, {Szegedi-Elek}, {Taris}, {Taylor}, {Teixeira}, {Tolomei}, {Tonello}, {Torra}, {Torra}, {Torralba Elipe}, {Trabucchi}, {Tsounis}, {Turon}, {Ulla}, {Unger}, {Vaillant}, {van Dillen}, {van Reeven}, {Vanel}, {Vecchiato}, {Viala}, {Vicente}, {Voutsinas}, {Weiler}, {Wevers}, {Wyrzykowski}, {Yoldas}, {Yvard}, {Zhao}, {Zorec}, {Zucker}, \& {Zwitter}}]{dr323}
{Gaia Collaboration}, {Vallenari}, A., {Brown}, A.~G.~A., {et~al.} 2023{\natexlab{b}}, \aap, 674, A1

\bibitem[{{Ge} {et~al.}(2024){Ge}, {Li}, {Hao}, {Lin}, {Hou}, {Liu}, {Li}, \& {Bian}}]{glh+24}
{Ge}, Q.~A., {Li}, J.~J., {Hao}, C.~J., {et~al.} 2024, \aj, 168, 25

\bibitem[{{Georgelin} \& {Georgelin}(1976)}]{gg76}
{Georgelin}, Y.~M. \& {Georgelin}, Y.~P. 1976, \aap, 49, 57

\bibitem[{{Hao} {et~al.}(2020){Hao}, {Xu}, {Wu}, {He}, \& {Bian}}]{hxw20}
{Hao}, C., {Xu}, Y., {Wu}, Z., {He}, Z., \& {Bian}, S. 2020, \pasp, 132, 034502

\bibitem[{{Hao} {et~al.}(2021){Hao}, {Xu}, {Hou}, {Bian}, {Li}, {Wu}, {He}, {Li}, \& {Liu}}]{hxh21}
{Hao}, C.~J., {Xu}, Y., {Hou}, L.~G., {et~al.} 2021, \aap, 652, A102

\bibitem[{{Hao} {et~al.}(2022){Hao}, {Xu}, {Wu}, {Lin}, {Liu}, \& {Li}}]{hxw22}
{Hao}, C.~J., {Xu}, Y., {Wu}, Z.~Y., {et~al.} 2022, \aap, 660, A4

\bibitem[{{Herbst} {et~al.}(1982){Herbst}, {Miller}, {Warner}, \& {Herzog}}]{hmwh82}
{Herbst}, W., {Miller}, D.~P., {Warner}, J.~W., \& {Herzog}, A. 1982, \aj, 87, 98

\bibitem[{{Honma} {et~al.}(2012){Honma}, {Nagayama}, {Ando}, {Bushimata}, {Choi}, {Handa}, {Hirota}, {Imai}, {Jike}, {Kim}, {Kameya}, {Kawaguchi}, {Kobayashi}, {Kurayama}, {Kuji}, {Matsumoto}, {Manabe}, {Miyaji}, {Motogi}, {Nakagawa}, {Nakanishi}, {Niinuma}, {Oh}, {Omodaka}, {Oyama}, {Sakai}, {Sato}, {Sato}, {Shibata}, {Shiozaki}, {Sunada}, {Tamura}, {Ueno}, \& {Yamauchi}}]{honm12}
{Honma}, M., {Nagayama}, T., {Ando}, K., {et~al.} 2012, \pasj, 64, 136

\bibitem[{{Hou} {et~al.}(2022){Hou}, {Han}, {Hong}, {Gao}, \& {Wang}}]{hhh22}
{Hou}, L., {Han}, J., {Hong}, T., {Gao}, X., \& {Wang}, C. 2022, Science China Physics, Mechanics, and Astronomy, 65, 129703

\bibitem[{{Hou}(2021)}]{hou21}
{Hou}, L.~G. 2021, Frontiers in Astronomy and Space Sciences, 8, 103

\bibitem[{{Hou} \& {Han}(2014)}]{hh14}
{Hou}, L.~G. \& {Han}, J.~L. 2014, \aap, 569, A125

\bibitem[{{Hou} \& {Han}(2015)}]{hh15}
{Hou}, L.~G. \& {Han}, J.~L. 2015, \mnras, 454, 626

\bibitem[{{Hou} {et~al.}(2009){Hou}, {Han}, \& {Shi}}]{hhs09}
{Hou}, L.~G., {Han}, J.~L., \& {Shi}, W.~B. 2009, \aap, 499, 473

\bibitem[{{Jing} {et~al.}(2023){Jing}, {Han}, {Hong}, {Wang}, {Gao}, {Hou}, {Zhou}, {Xu}, \& {Yang}}]{jhh+23}
{Jing}, W.~C., {Han}, J.~L., {Hong}, T., {et~al.} 2023, \mnras, 523, 4949

\bibitem[{{Joshi} \& {Malhotra}(2023)}]{jm23}
{Joshi}, Y.~C. \& {Malhotra}, S. 2023, \aj, 166, 170

\bibitem[{{Khoperskov} {et~al.}(2020){Khoperskov}, {Gerhard}, {Di Matteo}, {Haywood}, {Katz}, {Khrapov}, {Khoperskov}, \& {Arnaboldi}}]{kgd+20}
{Khoperskov}, S., {Gerhard}, O., {Di Matteo}, P., {et~al.} 2020, \aap, 634, L8

\bibitem[{{Klessen} \& {Glover}(2016)}]{kg16}
{Klessen}, R.~S. \& {Glover}, S. C.~O. 2016, Saas-Fee Advanced Course, 43, 85

\bibitem[{{Konietzka} {et~al.}(2024){Konietzka}, {Goodman}, {Zucker}, {Burkert}, {Alves}, {Foley}, {Swiggum}, {Koller}, \& {Miret-Roig}}]{kgz+24}
{Konietzka}, R., {Goodman}, A.~A., {Zucker}, C., {et~al.} 2024, \nat, 628, 62

\bibitem[{{Kuhn} {et~al.}(2021){Kuhn}, {Benjamin}, {Zucker}, {Krone-Martins}, {de Souza}, {Castro-Ginard}, {Ishida}, {Povich}, \& {Hillenbrand}}]{kbz+21}
{Kuhn}, M.~A., {Benjamin}, R.~A., {Zucker}, C., {et~al.} 2021, \aap, 651, L10

\bibitem[{{Lemasle} {et~al.}(2022){Lemasle}, {Lala}, {Kovtyukh}, {Hanke}, {Prudil}, {Bono}, {Braga}, {da Silva}, {Fabrizio}, {Fiorentino}, {Fran{\c{c}}ois}, {Grebel}, \& {Kniazev}}]{llk+22}
{Lemasle}, B., {Lala}, H.~N., {Kovtyukh}, V., {et~al.} 2022, \aap, 668, A40

\bibitem[{{Lequeux}(2005)}]{leq05}
{Lequeux}, J. 2005, {The Interstellar Medium}

\bibitem[{{Levine} {et~al.}(2006){Levine}, {Blitz}, \& {Heiles}}]{lev06}
{Levine}, E.~S., {Blitz}, L., \& {Heiles}, C. 2006, Science, 312, 1773

\bibitem[{{Li} \& {Chen}(2022)}]{lc22}
{Li}, G.-X. \& {Chen}, B.-Q. 2022, \mnras, 517, L102

\bibitem[{{Lin} \& {Shu}(1964)}]{lin+64}
{Lin}, C.~C. \& {Shu}, F.~H. 1964, \apj, 140, 646

\bibitem[{{Lin} \& {Shu}(1966)}]{lin+66}
{Lin}, C.~C. \& {Shu}, F.~H. 1966, Proceedings of the National Academy of Science, 55, 229

\bibitem[{{Lin} {et~al.}(2022){Lin}, {Xu}, {Hou}, {Liu}, {Li}, {Hao}, {Li}, \& {Bian}}]{lxh22}
{Lin}, Z., {Xu}, Y., {Hou}, L., {et~al.} 2022, \apj, 931, 72

\bibitem[{{Liu} {et~al.}(2019{\natexlab{a}}){Liu}, {Anderson}, {McIntyre}, {Anish Roshi}, {Churchwell}, {Minchin}, \& {Terzian}}]{siggma}
{Liu}, B., {Anderson}, L.~D., {McIntyre}, T., {et~al.} 2019{\natexlab{a}}, \apjs, 240, 14

\bibitem[{{Liu} {et~al.}(2016){Liu}, {Li}, {Wu}, {Yuan}, {Liu}, {Dubner}, {Paron}, {Ortega}, {Molinari}, {Huang}, {Zavagno}, {Samal}, {Huang}, \& {Zhang}}]{LiuHL16}
{Liu}, H.-L., {Li}, J.-Z., {Wu}, Y., {et~al.} 2016, \apj, 818, 95

\bibitem[{{Liu} {et~al.}(2015){Liu}, {Wu}, {Li}, {Yuan}, {Liu}, \& {Dong}}]{LiuHL15}
{Liu}, H.-L., {Wu}, Y., {Li}, J., {et~al.} 2015, \apj, 798, 30

\bibitem[{{Liu} \& {Pang}(2019)}]{lp19}
{Liu}, L. \& {Pang}, X. 2019, \apjs, 245, 32

\bibitem[{{Liu} {et~al.}(2019{\natexlab{b}}){Liu}, {Cui}, {Liu}, {Huang}, {Zhao}, \& {Zhang}}]{lcl+19}
{Liu}, Z., {Cui}, W., {Liu}, C., {et~al.} 2019{\natexlab{b}}, \apjs, 241, 32

\bibitem[{{Lockman}(1989)}]{lockman89}
{Lockman}, F.~J. 1989, \apjs, 71, 469

\bibitem[{{Maciel} \& {Costa}(2010)}]{mc10}
{Maciel}, W.~J. \& {Costa}, R. D.~D. 2010, in Chemical Abundances in the Universe: Connecting First Stars to Planets, ed. K.~{Cunha}, M.~{Spite}, \& B.~{Barbuy}, Vol. 265, 317--324

\bibitem[{{Ma{\'\i}z Apell{\'a}niz} {et~al.}(2016){Ma{\'\i}z Apell{\'a}niz}, {Sota}, {Arias}, {Barb{\'a}}, {Walborn}, {Sim{\'o}n-D{\'\i}az}, {Negueruela}, {Marco}, {Le{\~a}o}, {Herrero}, {Gamen}, \& {Alfaro}}]{msa+16}
{Ma{\'\i}z Apell{\'a}niz}, J., {Sota}, A., {Arias}, J.~I., {et~al.} 2016, \apjs, 224, 4

\bibitem[{{Martinez-Medina} {et~al.}(2022){Martinez-Medina}, {P{\'e}rez-Villegas}, \& {Peimbert}}]{mpp22}
{Martinez-Medina}, L., {P{\'e}rez-Villegas}, A., \& {Peimbert}, A. 2022, \mnras, 512, 1574

\bibitem[{{McClure-Griffiths} {et~al.}(2005){McClure-Griffiths}, {Dickey}, {Gaensler}, {Green}, {Haverkorn}, \& {Strasser}}]{sgps}
{McClure-Griffiths}, N.~M., {Dickey}, J.~M., {Gaensler}, B.~M., {et~al.} 2005, \apjs, 158, 178

\bibitem[{{McKee} \& {Williams}(1997)}]{mw97}
{McKee}, C.~F. \& {Williams}, J.~P. 1997, \apj, 476, 144

\bibitem[{{M{\'e}ndez-Delgado} {et~al.}(2022){M{\'e}ndez-Delgado}, {Amayo}, {Arellano-C{\'o}rdova}, {Esteban}, {Garc{\'\i}a-Rojas}, {Carigi}, \& {Delgado-Inglada}}]{maa+22}
{M{\'e}ndez-Delgado}, J.~E., {Amayo}, A., {Arellano-C{\'o}rdova}, K.~Z., {et~al.} 2022, \mnras, 510, 4436

\bibitem[{{Miyachi} {et~al.}(2019){Miyachi}, {Sakai}, {Kawata}, {Baba}, {Honma}, {Matsunaga}, \& {Fujisawa}}]{msk+19}
{Miyachi}, Y., {Sakai}, N., {Kawata}, D., {et~al.} 2019, \apj, 882, 48

\bibitem[{{Moffat} {et~al.}(1979){Moffat}, {Fitzgerald}, \& {Jackson}}]{mfg79}
{Moffat}, A.~F.~J., {Fitzgerald}, M.~P., \& {Jackson}, P.~D. 1979, \aaps, 38, 197

\bibitem[{{Mohr-Smith} {et~al.}(2015){Mohr-Smith}, {Drew}, {Barentsen}, {Wright}, {Napiwotzki}, {Corradi}, {Eisl{\"o}ffel}, {Groot}, {Kalari}, {Parker}, {Raddi}, {Sale}, {Unruh}, {Vink}, \& {Wesson}}]{mdb+15}
{Mohr-Smith}, M., {Drew}, J.~E., {Barentsen}, G., {et~al.} 2015, \mnras, 450, 3855

\bibitem[{{Mohr-Smith} {et~al.}(2017){Mohr-Smith}, {Drew}, {Napiwotzki}, {Sim{\'o}n-D{\'\i}az}, {Wright}, {Barentsen}, {Eisl{\"o}ffel}, {Farnhill}, {Greimel}, {Mongui{\'o}}, {Kalari}, {Parker}, \& {Vink}}]{mdn+17}
{Mohr-Smith}, M., {Drew}, J.~E., {Napiwotzki}, R., {et~al.} 2017, \mnras, 465, 1807

\bibitem[{{Mois{\'e}s} {et~al.}(2011){Mois{\'e}s}, {Damineli}, {Figuer{\^e}do}, {Blum}, {Conti}, \& {Barbosa}}]{mdf+11}
{Mois{\'e}s}, A.~P., {Damineli}, A., {Figuer{\^e}do}, E., {et~al.} 2011, \mnras, 411, 705

\bibitem[{{Mongui{\'o}} {et~al.}(2015){Mongui{\'o}}, {Grosb{\o}l}, \& {Figueras}}]{mgf15}
{Mongui{\'o}}, M., {Grosb{\o}l}, P., \& {Figueras}, F. 2015, \aap, 577, A142

\bibitem[{{Morgan} {et~al.}(1953){Morgan}, {Whitford}, \& {Code}}]{mwc53}
{Morgan}, W.~W., {Whitford}, A.~E., \& {Code}, A.~D. 1953, \apj, 118, 318

\bibitem[{{Oort} \& {Muller}(1952)}]{oort52}
{Oort}, J.~H. \& {Muller}, C.~A. 1952, Monthly Notes of the Astronomical Society of South Africa, 11, 65

\bibitem[{{Paladini} {et~al.}(2004){Paladini}, {Davies}, \& {De Zotti}}]{pdd04}
{Paladini}, R., {Davies}, R.~D., \& {De Zotti}, G. 2004, \mnras, 347, 237

\bibitem[{{Pantaleoni Gonz{\'a}lez} {et~al.}(2021){Pantaleoni Gonz{\'a}lez}, {Ma{\'\i}z Apell{\'a}niz}, {Barb{\'a}}, \& {Reed}}]{pmb+21}
{Pantaleoni Gonz{\'a}lez}, M., {Ma{\'\i}z Apell{\'a}niz}, J., {Barb{\'a}}, R.~H., \& {Reed}, B.~C. 2021, \mnras, 504, 2968

\bibitem[{{Poggio} {et~al.}(2021){Poggio}, {Drimmel}, {Cantat-Gaudin}, {Ramos}, {Ripepi}, {Zari}, {Andrae}, {Blomme}, {Chemin}, {Clementini}, {Figueras}, {Fouesneau}, {Fr{\'e}mat}, {Lobel}, {Marshall}, {Muraveva}, \& {Romero-G{\'o}mez}}]{pdc+21}
{Poggio}, E., {Drimmel}, R., {Cantat-Gaudin}, T., {et~al.} 2021, \aap, 651, A104

\bibitem[{{Quireza} {et~al.}(2006){Quireza}, {Rood}, {Bania}, {Balser}, \& {Maciel}}]{qrbb+06}
{Quireza}, C., {Rood}, R.~T., {Bania}, T.~M., {Balser}, D.~S., \& {Maciel}, W.~J. 2006, \apj, 653, 1226

\bibitem[{{Reed}(2003)}]{reed03}
{Reed}, B.~C. 2003, \aj, 125, 2531

\bibitem[{{Reid} {et~al.}(2019){Reid}, {Menten}, {Brunthaler}, {Zheng}, {Dame}, {Xu}, {Li}, {Sakai}, {Wu}, {Immer}, {Zhang}, {Sanna}, {Moscadelli}, {Rygl}, {Bartkiewicz}, {Hu}, {Quiroga-Nu{\~n}ez}, \& {van Langevelde}}]{reid19}
{Reid}, M.~J., {Menten}, K.~M., {Brunthaler}, A., {et~al.} 2019, \apj, 885, 131

\bibitem[{{Reid} {et~al.}(2014){Reid}, {Menten}, {Brunthaler}, {Zheng}, {Dame}, {Xu}, {Wu}, {Zhang}, {Sanna}, {Sato}, {Hachisuka}, {Choi}, {Immer}, {Moscadelli}, {Rygl}, \& {Bartkiewicz}}]{reid14}
{Reid}, M.~J., {Menten}, K.~M., {Brunthaler}, A., {et~al.} 2014, \apj, 783, 130

\bibitem[{{Reid} {et~al.}(2009){Reid}, {Menten}, {Zheng}, {Brunthaler}, {Moscadelli}, {Xu}, {Zhang}, {Sato}, {Honma}, {Hirota}, {Hachisuka}, {Choi}, {Moellenbrock}, \& {Bartkiewicz}}]{rmz+09}
{Reid}, M.~J., {Menten}, K.~M., {Zheng}, X.~W., {et~al.} 2009, \apj, 700, 137

\bibitem[{{Roberts}(1969)}]{rob69}
{Roberts}, W.~W. 1969, \apj, 158, 123

\bibitem[{{Roslund}(1963)}]{roslund63}
{Roslund}, C. 1963, Arkiv for Astronomi, 3, 97

\bibitem[{{Rudolph} {et~al.}(2006){Rudolph}, {Fich}, {Bell}, {Norsen}, {Simpson}, {Haas}, \& {Erickson}}]{rfb+06}
{Rudolph}, A.~L., {Fich}, M., {Bell}, G.~R., {et~al.} 2006, \apjs, 162, 346

\bibitem[{{Russeil}(2003)}]{rus03}
{Russeil}, D. 2003, \aap, 397, 133

\bibitem[{{Russeil} {et~al.}(2007){Russeil}, {Adami}, \& {Georgelin}}]{rag07}
{Russeil}, D., {Adami}, C., \& {Georgelin}, Y.~M. 2007, \aap, 470, 161

\bibitem[{{Sellwood} \& {Carlberg}(1984)}]{sc84}
{Sellwood}, J.~A. \& {Carlberg}, R.~G. 1984, \apj, 282, 61

\bibitem[{{Sewilo} {et~al.}(2004){Sewilo}, {Churchwell}, {Kurtz}, {Goss}, \& {Hofner}}]{sck+04}
{Sewilo}, M., {Churchwell}, E., {Kurtz}, S., {Goss}, W.~M., \& {Hofner}, P. 2004, \apj, 605, 285

\bibitem[{{Shen} \& {Zheng}(2020)}]{sz20}
{Shen}, J. \& {Zheng}, X.-W. 2020, Research in Astronomy and Astrophysics, 20, 159

\bibitem[{{Shu}(2016)}]{shu+16}
{Shu}, F.~H. 2016, \araa, 54, 667

\bibitem[{{Skiff}(2014)}]{skiff14}
{Skiff}, B.~A. 2014, {VizieR Online Data Catalog: Catalogue of Stellar Spectral Classifications (Skiff, 2009- )}, VizieR On-line Data Catalog: B/mk. Originally published in: Lowell Observatory (October 2014)

\bibitem[{{Smith} \& {Kennicutt}(1989)}]{sk89}
{Smith}, T.~R. \& {Kennicutt}, Robert~C., J. 1989, \pasp, 101, 649

\bibitem[{{Soubiran} {et~al.}(2018){Soubiran}, {Cantat-Gaudin}, {Romero-G{\'o}mez}, {Casamiquela}, {Jordi}, {Vallenari}, {Antoja}, {Balaguer-N{\'u}{\~n}ez}, {Bossini}, {Bragaglia}, {Carrera}, {Castro-Ginard}, {Figueras}, {Heiter}, {Katz}, {Krone-Martins}, {Le Campion}, {Moitinho}, \& {Sordo}}]{scr18}
{Soubiran}, C., {Cantat-Gaudin}, T., {Romero-G{\'o}mez}, M., {et~al.} 2018, \aap, 619, A155

\bibitem[{{Stil} {et~al.}(2006){Stil}, {Taylor}, {Dickey}, {Kavars}, {Martin}, {Rothwell}, {Boothroyd}, {Lockman}, \& {McClure-Griffiths}}]{vgps}
{Stil}, J.~M., {Taylor}, A.~R., {Dickey}, J.~M., {et~al.} 2006, \aj, 132, 1158

\bibitem[{{Swiggum} {et~al.}(2022){Swiggum}, {Alves}, {D'Onghia}, {Benjamin}, {Thulasidharan}, {Zucker}, {Poggio}, {Drimmel}, {Gallagher}, \& {Goodman}}]{sad+2022}
{Swiggum}, C., {Alves}, J., {D'Onghia}, E., {et~al.} 2022, \aap, 664, L13

\bibitem[{{Taylor} {et~al.}(2003){Taylor}, {Gibson}, {Peracaula}, {Martin}, {Landecker}, {Brunt}, {Dewdney}, {Dougherty}, {Gray}, {Higgs}, {Kerton}, {Knee}, {Kothes}, {Purton}, {Uyaniker}, {Wallace}, {Willis}, \& {Durand}}]{cgps}
{Taylor}, A.~R., {Gibson}, S.~J., {Peracaula}, M., {et~al.} 2003, \aj, 125, 3145

\bibitem[{{Thulasidharan} {et~al.}(2022){Thulasidharan}, {D'Onghia}, {Poggio}, {Drimmel}, {Gallagher}, {Swiggum}, {Benjamin}, \& {Alves}}]{tdp+2022}
{Thulasidharan}, L., {D'Onghia}, E., {Poggio}, E., {et~al.} 2022, \aap, 660, L12

\bibitem[{{Toomre} \& {Toomre}(1972)}]{tt72}
{Toomre}, A. \& {Toomre}, J. 1972, \apj, 178, 623

\bibitem[{{Uppal} {et~al.}(2023){Uppal}, {Ganesh}, \& {Schultheis}}]{ugs23}
{Uppal}, N., {Ganesh}, S., \& {Schultheis}, M. 2023, \aap, 673, A99

\bibitem[{{Vall{\'e}e}(2014)}]{val14}
{Vall{\'e}e}, J.~P. 2014, \aj, 148, 5

\bibitem[{{Vall{\'e}e}(2022)}]{vall22}
{Vall{\'e}e}, J.~P. 2022, \apss, 367, 26

\bibitem[{{van de Hulst} {et~al.}(1954){van de Hulst}, {Muller}, \& {Oort}}]{vmo54}
{van de Hulst}, H.~C., {Muller}, C.~A., \& {Oort}, J.~H. 1954, \bain, 12, 117

\bibitem[{{V{\'a}zquez} {et~al.}(2008){V{\'a}zquez}, {May}, {Carraro}, {Bronfman}, {Moitinho}, \& {Baume}}]{vmc+08}
{V{\'a}zquez}, R.~A., {May}, J., {Carraro}, G., {et~al.} 2008, \apj, 672, 930

\bibitem[{{VERA Collaboration} {et~al.}(2020){VERA Collaboration}, {Hirota}, {Nagayama}, {Honma}, {Adachi}, {Burns}, {Chibueze}, {Choi}, {Hachisuka}, {Hada}, {Hagiwara}, {Hamada}, {Handa}, {Hashimoto}, {Hirano}, {Hirata}, {Ichikawa}, {Imai}, {Inenaga}, {Ishikawa}, {Jike}, {Kameya}, {Kaseda}, {Kim}, {Kim}, {Kim}, {Kobayashi}, {Kono}, {Kurayama}, {Matsuno}, {Morita}, {Motogi}, {Murase}, {Nakagawa}, {Nakanishi}, {Niinuma}, {Nishi}, {Oh}, {Omodaka}, {Oyadomari}, {Oyama}, {Sakai}, {Sakai}, {Sawada-Satoh}, {Shibata}, {Shizugami}, {Sudo}, {Sugiyama}, {Sunada}, {Suzuki}, {Takahashi}, {Tamura}, {Tazaki}, {Ueno}, {Uno}, {Urago}, {Wada}, {Wu}, {Yamashita}, {Yamashita}, {Yamauchi}, \& {Yuda}}]{vera20}
{VERA Collaboration}, {Hirota}, T., {Nagayama}, T., {et~al.} 2020, \pasj, 72, 50

\bibitem[{{Wenger} {et~al.}(2000){Wenger}, {Ochsenbein}, {Egret}, {Dubois}, {Bonnarel}, {Borde}, {Genova}, {Jasniewicz}, {Lalo{\"e}}, {Lesteven}, \& {Monier}}]{simbad}
{Wenger}, M., {Ochsenbein}, F., {Egret}, D., {et~al.} 2000, \aaps, 143, 9

\bibitem[{{Wenger} {et~al.}(2018){Wenger}, {Balser}, {Anderson}, \& {Bania}}]{wba+18}
{Wenger}, T.~V., {Balser}, D.~S., {Anderson}, L.~D., \& {Bania}, T.~M. 2018, \apj, 856, 52

\bibitem[{{Wenger} {et~al.}(2019){Wenger}, {Balser}, {Anderson}, \& {Bania}}]{wba+19}
{Wenger}, T.~V., {Balser}, D.~S., {Anderson}, L.~D., \& {Bania}, T.~M. 2019, \apj, 887, 114

\bibitem[{{Wenger} {et~al.}(2021){Wenger}, {Dawson}, {Dickey}, {Jordan}, {McClure-Griffiths}, {Anderson}, {Armentrout}, {Balser}, \& {Bania}}]{shrds21}
{Wenger}, T.~V., {Dawson}, J.~R., {Dickey}, J.~M., {et~al.} 2021, \apjs, 254, 36

\bibitem[{{Widmark} \& {Naik}(2024)}]{wn24}
{Widmark}, A. \& {Naik}, A.~P. 2024, arXiv e-prints, arXiv:2401.04571

\bibitem[{{Wink} {et~al.}(1983){Wink}, {Wilson}, \& {Bieging}}]{wwb83}
{Wink}, J.~E., {Wilson}, T.~L., \& {Bieging}, J.~H. 1983, \aap, 127, 211

\bibitem[{{Wright} {et~al.}(2010){Wright}, {Eisenhardt}, {Mainzer}, {Ressler}, {Cutri}, {Jarrett}, {Kirkpatrick}, {Padgett}, {McMillan}, {Skrutskie}, {Stanford}, {Cohen}, {Walker}, {Mather}, {Leisawitz}, {Gautier}, {McLean}, {Benford}, {Lonsdale}, {Blain}, {Mendez}, {Irace}, {Duval}, {Liu}, {Royer}, {Heinrichsen}, {Howard}, {Shannon}, {Kendall}, {Walsh}, {Larsen}, {Cardon}, {Schick}, {Schwalm}, {Abid}, {Fabinsky}, {Naes}, \& {Tsai}}]{wise}
{Wright}, E.~L., {Eisenhardt}, P. R.~M., {Mainzer}, A.~K., {et~al.} 2010, \aj, 140, 1868

\bibitem[{{Xu} {et~al.}(2018{\natexlab{a}}){Xu}, {Bian}, {Reid}, {Li}, {Zhang}, {Yan}, {Dame}, {Menten}, {He}, {Liao}, \& {Tang}}]{xu+2018}
{Xu}, Y., {Bian}, S.~B., {Reid}, M.~J., {et~al.} 2018{\natexlab{a}}, \aap, 616, L15

\bibitem[{{Xu} {et~al.}(2023){Xu}, {Hao}, {Liu}, {Lin}, {Bian}, {Hou}, {Li}, \& {Li}}]{xu+23}
{Xu}, Y., {Hao}, C.~J., {Liu}, D.~J., {et~al.} 2023, \apj, 947, 54

\bibitem[{{Xu} {et~al.}(2021){Xu}, {Hou}, {Bian}, {Hao}, {Liu}, {Li}, \& {Li}}]{xhb+21}
{Xu}, Y., {Hou}, L.~G., {Bian}, S.~B., {et~al.} 2021, \aap, 645, L8

\bibitem[{{Xu} {et~al.}(2018{\natexlab{b}}){Xu}, {Hou}, \& {Wu}}]{xhw18}
{Xu}, Y., {Hou}, L.-G., \& {Wu}, Y.-W. 2018{\natexlab{b}}, Research in Astronomy and Astrophysics, 18, 146

\bibitem[{{Xu} {et~al.}(2013){Xu}, {Li}, {Reid}, {Menten}, {Zheng}, {Brunthaler}, {Moscadelli}, {Dame}, \& {Zhang}}]{xlr+13}
{Xu}, Y., {Li}, J.~J., {Reid}, M.~J., {et~al.} 2013, \apj, 769, 15

\bibitem[{{Xu} {et~al.}(2016){Xu}, {Reid}, {Dame}, {Menten}, {Sakai}, {Li}, {Brunthaler}, {Moscadelli}, {Zhang}, \& {Zheng}}]{xu+16}
{Xu}, Y., {Reid}, M., {Dame}, T., {et~al.} 2016, Science Advances, 2, e1600878

\bibitem[{{Xu} {et~al.}(2006){Xu}, {Reid}, {Zheng}, \& {Menten}}]{xrzm06}
{Xu}, Y., {Reid}, M.~J., {Zheng}, X.~W., \& {Menten}, K.~M. 2006, Science, 311, 54

\bibitem[{{Yu} \& {Ho}(2018)}]{yu+18}
{Yu}, S.-Y. \& {Ho}, L.~C. 2018, \apj, 869, 29

\bibitem[{{Zhu} {et~al.}(2024){Zhu}, {Fang}, {Lu}, {Wang}, {Li}, {Zhang}, {Pelkonen}, {Padoan}, \& {Liang}}]{zfl+24}
{Zhu}, Z.-K., {Fang}, M., {Lu}, Z.-J., {et~al.} 2024, \apj, 971, 167

\end{thebibliography}

\onecolumn
\begin{appendix}
\label{appendix}

\begin{landscape}
\section{Tables}
\label{appendix-tables}

\renewcommand{\arraystretch}{1.3}
\tabcolsep3pt 
\begin{longtable}{llcccccccccc}

\caption{Information of exciting 
 stars.\label{exciting stars}}\\
\hline
\hline
$\hii$ region & Stars & Spectral &  $l$ & $b$ & Ref. & Gaia DR3 ID & $\varpi$ & $\mu_{\alpha}$ & $\mu_{\delta}$ & d$_{\sun}$\\
& & Type& ($\degr$) & ($\degr$) & & &(mas) &(mas~yr$^{-1}$) &(mas~yr$^{-1}$)& (kpc)& \\
(1)&(2) & (3)& (4) & (5) &(6) & (7) &(8) &(9) &(10)& (11)\\
\hline
\endfirsthead
\hline
\endfoot
G000.003+00.127 & 1 & OB & 0.00902058 & 0.1416196 & Chen & 4057487597750725376 & 0.3425$\pm0.0511$ & $0.128\pm0.055$ & $-0.581\pm0.033$ & $2.63_{-0.32}^{+0.53}$\\
G000.003+00.127 & - & OB & 0.00592264 & 0.1296308 & Chen & 4057487529031245824 & 0.4595$\pm0.0019$ & $0.082\pm0.02$ & $-0.606\pm0.012$ & $1.99_{-0.09}^{+0.07}$\\
G000.003+00.127 & - & OB & 0.00615124 & 0.133549 & Chen & 4057487524729379200 & 0.3965$\pm0.021$ & $0.136\pm0.023$ & $-0.395\pm0.014$ & $2.20_{-0.10}^{+0.11}$\\
G000.121$-$00.304 & -& OB & 0.12175563 & -0.30479151 & Chen & 4057473407178966912 & 0.3509$\pm0.0182$ & $0.122\pm0.017$ & $-0.611\pm0.011$ & $2.51_{-0.11}^{+0.15}$\\
G006.664$-$01.545 & 2& B0.5Vn & 6.65730192 & -1.53822774 & Skiff$^a$ &  4066103267811109504 & 0.8120$\pm0.0227$ & $2.043\pm0.025$ & $-1.239\pm0.017$ & $1.19_{-0.03}^{+0.03}$\\
G017.315+00.389 & 3& OB & 17.31811908 & 0.39089841 & Chen &  4152422554127130240 & 0.5222$\pm0.0152$ & $-0.16\pm0.016$ & $-1.864\pm0.013$ & $1.85_{-0.05}^{+0.05}$ \\
G018.253$-$00.298 & 4&O6V-O5V & 18.26605645 & -0.29873344 & Simbad &  4152566800618412928 & 0.2418$\pm0.0178$ & $0.073\pm0.019$ & $-1.466\pm0.015$ & $3.43_{-0.22}^{+0.20}$\\
G018.669+01.965 & 5 & OB & 18.68589962 & 1.97021693 & Chen &  4153878479319865472 & 0.4599$\pm0.0242$ & $-0.392\pm0.025$ & $-1.711\pm0.021$ & $2.03_{-0.09}^{+0.12}$\\
G018.669+01.965 & - & OB & 18.66782964 & 1.96907581 & Chen &  4153878449263407616 & 0.4992$\pm0.0234$ & $-0.618\pm0.024$ & $-2.02\pm0.02$ & $1.87_{-0.08}^{+0.10}$\\
G024.138+00.123 & 6 & OB & 24.13550601 & 0.1153291 & Skiff$^b$ &  4156737621868942208 & 0.4292$\pm0.014$ & $0.108\pm0.015$ & $-0.308\pm0.013$ & $2.18_{-0.06}^{+0.08}$\\
...& ... & ... & ... & ...&...& ...& ...& ... & ... &...\\
\\
\hline
\end{longtable}
\tablefoot{The entire version in a machine-readable format is available at the CDS. Col.\,(1) is the $\hii$ region name given in \cite{abb+14}; Cols\,(2), (3), (4), (5) and  (6) are the exciting star name, spectral type, the coordinate of Galactic longitude, the coordinate of Galactic latitude and reference for star. Cols.\,(7), (8), (9) and (10) list the Gaia DR3 ID of matched source, Gaia parallax, proper motion in the RA direction and proper motion in the Dec direction. Col.\,(11) gives the adopted parallax distance .\\
{\bf Exciting star name:} 1:\,2MASS J17450537-2851166;
2:\,CD-23 13906;
3:\,UCAC4 382-102367; 4:\,2MASS J18251808-1309427; 5:\,GSC 05685-04088; 6:\,[R63] 11\\
{\bf Ref.}\\ 
Chen:\,\cite{chh+19};\\
Skiff:\,\cite{skiff14};\\
a:\,\cite{hmwh82}
b:\,\cite{roslund63}}

\end{landscape}

\end{appendix}
\end{document}